\documentclass[a4paper,11pt]{article}
\pdfoutput=1 
\usepackage{jcappub} 
\usepackage[T1]{fontenc} 
\usepackage[utf8x]{inputenc}
\usepackage{multirow}
\usepackage{booktabs}
\usepackage{bm}
\usepackage{times}
\usepackage{tensor}
\title{Strange Stars within Bosonic and Fermionic Admixed Dark Matter}
\author[a]{Luiz L. Lopes,}
\author[b,1]{H. C. Das \note{Corresponding author.}}

\affiliation[a]{Centro Federal de Educa\c{c}\~ao Tecnol\'ogica de Minas Gerais Campus VIII; CEP 37.022-560, Varginha - MG - Brasil.}
\affiliation[b]{Institute of Physics, Sachivalaya Marg, Bhubaneswar 751005, India.}
\emailAdd{llopes@cefetmg.br}
\emailAdd{harish.d@iopb.res.in}
\abstract{
In this work, we study dark matter (DM) admixed strange quark stars exploring the different possibilities about the nature of the DM and their effects on the macroscopic properties of strange stars, such as maximum masses, radii, as well the dimensionless tidal parameter. We observe that the DM significantly affects the macroscopic properties that depend on its mass, type, and fraction inside the star.
}
\begin{document}
\maketitle
\flushbottom
\section{Introduction}
Recently, it was suggested that quarks probably appear inside the core of a massive neutron star (NS) due to a very high density, where hadronic matter undergoes phase transitions to a new phase of quarks and gluons \cite{Annala_2020}. Furthermore, several exotic particles, such as hyperons production, dark matter accretions, kaon condensations, and so on, appeared primarily in the core of the NS. As a result, in those regimes, the equation of state (EoS) is the fundamental component that can describe both micro/macroscopic properties. Another possibility is that at least some of the observed pulsars are indeed stable quark stars or strange stars. Several theoretical works have proposed the existence of strange quark stars (SQSs)~\cite{xu_2003}, which are made up of $u$, $d$, and $s$ quarks in equilibrium in terms of weak interactions. The Bodmen-Witten conjecture states that strange quark matter (SQM) can have a lower energy per baryon than pure nucleons because the exclusion principle may be dominant at absolute zero pressure and temperature \cite{Bodmer_1971, Witten_1984}. Hence, the SQM might be the true ground state of the hadronic matter. Hence, it stands to reason that the SQS must be more stable than the ordinary NS. 

Various phenomenological models have been proposed to explore the SQSs properties. Among them, the MIT bag model and Nambu-Jona-Lasinio (NJL) have been widely used. In this study, we use the vector MIT (vMIT) bag model to describe the quark matter interactions \cite{Lopes_2021, Lopes_2022}. In the MIT bag model, it has been assumed that the quarks are bound in a bag of finite dimensions. In contrast to their absolute mass, which is very high, it is hypothesized that quarks inside the bag have a very low mass. The system is given a bag constant $B$ as a constant energy density to balance the bag's behavior and determine its size. The inward pressure at the bag's surface counterbalances the outward pressure the quarks cause, which means the pressure between the true and perturbative vacuum. Consequently, as $B$ increases, the quark pressure lowers, which impacts the star's structure. The value of $B$ relies on the mass of the strange quark when $u$ and $d$ quarks have very low masses. The values of $B$ still need to be established and are fully model-dependent. One can constrain its values with the help of observational results. For example, in the observational limit of GW170817, the predicted values of $B^{1/4}=134.1-141.4$ MeV with low-spin prior and $B^{1/4} = 126.1 - 141.4$ MeV with high spin prior for SQSs \cite{Zhou_2018}. In Ref. \cite{Aziz_2019}, they have predicted the range of $B^{1/4}=133.68 - 222.5$ MeV for SQSs. However, in the vMIT bag model \cite{Lopes_2021, Lopes_2022}, the value of $B$ can be obtained by including the stability window, as mentioned in Refs. \cite{Bodmer_1971, Witten_1984}.

Moreover, there are different phenomenological and macroscopic studies  suggesting that the quark phases inside the compact stars can undergo a phase transition into a color superconducting state of 2-flavour superconducting (2SC), and color-flavor locked (CFL) \cite{Alford_2004, Shovkovy_2005}. They form Cooper pairs at high density and low temperature \cite{Alford2008}. The gap parameter ($\Delta$) determines the pairing strength of Cooper pairs influence the formation of pure CFL stars \cite{Alford_2003, Drago_2004, Flores_2017, Lopes_2019, Bogadi_2020, Li_A_2021} and CFL magnetars \cite{Wen_2013, Liang_2019}. Recently, it has been suggested that with the proper choice of $\Delta$ and bag pressure $B$, the CFL stars and their EoS can successfully reproduce various observational constraints such as GW and NICER results \cite{Roupas_2014, Miao_2021, Lourenco_2021}. In this study, we want to explore the dark matter (DM) effects on the strange stars with and without CFL phases and try to constrain the macroscopic properties with various observational data. 

Compact objects such as NS, and white dwarfs may capture some amount of DM inside them in their evolving time due to their immense gravitational potential. Various theoretical predictions provide us with the unknown nature of DM. Still, numerous work has been fully dedicated to explaining its properties by applying it to different systems such as white dwarf \cite{Bell_2021}, NS \cite{Das:2020, Das_2021, Das:2021yny, Panotopoulos_2017, Guha_2021}, and even our earth \cite{GREEN2019120}. In the present study, we assume that the SQSs might contain a certain amount of DM in their lifetime. The types of DM particles may be either bosonic or fermionic, and also the percentage of DM depends on the (i) evolution time and (ii) types of accretions. However, the accreted DM particles interact directly or indirectly with hadrons by exchanging other bosonic particles, mainly depending on the model used. Here, we take different types of possible scenarios for DM admixed SQS. 

The direct detection experiments have already been established, such as XENON100 \cite{Aprile_2012}, XENON1T\cite{Aprile_2018}, CDMS \cite{Agnese_2018}, LUX \cite{Akerib_2013}, PANDAX-II \cite{Wang_2020} etc. to measure the scattering cross-section of the DM and nucleons. Although, they provided some exclusions bound to the scattering cross-section. Still, the null results provided by the experiments alluded to an inconclusive nature of DM. However, the exclusion bounds prescribed by such direct detection experiments depend on the local DM density around the solar neighborhood, which does not affect the density of DM in the NS/SQS environment. After the accretion of DM inside NS/SQS, it collides with nucleons or quarks by losing its kinetic energy, and eventually, it is bound inside the star. When the accretion ends, the DM particles finally reach thermal equilibrium with one another due to their internal interactions. This explains why NSs with admixed DM have essentially constant DM particle densities \cite{Panotopoulos_2017, Das_2021, OdilonDM, Guha_2021}. Therefore, the accreted DM particles are restricted to a narrow radius area inside the star. In this study, we choose two types of DM and see their effects on the SQS properties with the vMIT bag model and a model with superconducting phases. 

Recently, the fastest and heaviest Galactic NS named PSR J0952-0607 (black widow) in the disk of the Milky Way has been detected to have mass $M=2.35\pm0.17 \ M_\odot$ in continuation of the pulsars PSR J0740+6620 ($M=2.08\pm0.07 \ M_\odot$ ~\cite{Cromartie_2020, Fonseca_2021}). The simultaneous measurements of the $M$ and $R$ for NS are done by neutron star interior composition explorer (NICER) \citep{Miller_2019,Riley_2019} while the limit on the dimensionless tidal deformability of $\Lambda_{1.4}=190_{-120}^{+390}$ was provided  in GW170817 event \cite{Abbott_2018}. We calculate the mass, radius, and tidal deformability for the DM admixed SQS and put constraints using the observational results obtained from different x-ray/pulsars data, GW170817 data. 
\section{Formalism}
\subsection{Vector MIT bag model}
We use the thermodynamic consistent vector MIT bag model introduced in Ref.~\cite{Lopes_2021, Lopes_2022} to describe the quark matter. In this model, the quark interaction is mediated by the vector channel $V^\mu$, analogous to the $\omega$ meson in QHD~\cite{Serot_1992}. Its Lagrangian reads:
\begin{eqnarray}
\mathcal{L}_{\rm vMIT} &=& \bigg\{ \bar{\psi}_q\big[\gamma^\mu(i\partial_\mu - g_{qV} V_\mu) - m_q\big]\psi_q  \nonumber \\
&&
- B + \frac{1}{2}m_V^2V^\mu V_\mu  \bigg\}\Theta(\bar{\psi}_q\psi_q) ,
\label{vMIT}
\end{eqnarray}
where $m_q$ is the mass of the quark $q$ of flavor $u$, $d$ or $s$, $\psi_q$ is the Dirac quark fleld, $B$ is the constant vacuum pressure, and $\Theta(\bar{\psi}_q\psi_q)$ is the Heaviside step function to assure that the quarks exist only conflned to the bag. Applying Euler-Lagrange, we obtain the energy eigenvalue ($E_q$), which at $T = 0$ K, is also the chemical potential ($\mu_q$):
\begin{equation}
E_q = \mu_q = \sqrt{m_q^2 + k^2} + g_{qV}V_\mu,
\label{E}
\end{equation}
now, using Fermi-Dirac statistics, we can obtain the EoS in mean field approximation. The number density ($n_q)$ and the  energy density ($\epsilon_q$) of the quarks are:
\begin{eqnarray}
n_q  =  \frac{N_c}{3\pi^2}k_f^3, \nonumber \\
\epsilon_q = \frac{N_c}{\pi^2}\int_0^{k_f} E_q k^2 d^3k , \label{ed}
\end{eqnarray}
where $N_c = 3$ is the number of colors and $k_f$ is the Fermi momentum. The contribution of the bag and the mesonic mass term is obtained with the Hamiltonian: $\mathcal{H}$ = $- \langle \mathcal{L} \rangle$. The total number density ($n$) and energy density ($\epsilon$) of the quarks read:
\begin{eqnarray}
n = \sum_q n_q, \nonumber \\ 
\epsilon = \sum_q\epsilon_q + B - \frac{1}{2}m_v^2V_0^2. \label{ted}
\end{eqnarray}
To construct an electrically neutral, beta-stable matter, leptons are added as a free Fermi gas. The pressure is obtained via the relation: $p = \sum \mu n - \epsilon$, where the sum runs over all the fermions.

The parameters utilized in this work are the same as presented in Ref.~\cite{Lopes_2021}. We use $m_u = m_d$ = 4 MeV, and $m_s$ = 95 MeV. We also assume a universal coupling of quarks with the vector meson, i.e., $g_{uV} = g_{dV} = g_{sV} = g_V$, and use a value of $G_V$ = 0.3 fm$^{2}$ as defined below:
\begin{equation}
G_V = \bigg ( \frac{g_V}{m_V} \bigg )^2 = 0.3 \; \mbox{fm}^{2}.
\label{GV}
\end{equation}
Now, the value of $G_V$ is somewhat arbitrary. To reproduce stable strange matter, the value of $G_V$ combined with the bag must lie in the range known as the stability window. The stability window is related to the so-called Bodmer-Witten conjecture~\cite{Bodmer_1971, Witten_1984}, which states that the true ground state of the strongly interacting matter is not protons and neutrons but consists of strange quark matter, which in turn is composed of deconfined up, down, and strange quarks. For the SQM hypothesis to be accurate, the energy per baryon of the deconfined phase (for $p = 0$ and $T = 0$) is lower than the nonstrange infinite baryonic matter~\cite{Lopes_2021, Bodmer_1971, Witten_1984}.
\begin{equation}
E_{uds}/A < 930 \; \mbox{MeV},
\label{lower930}
\end{equation}
at the same time, the nonstrange matter still needs to have an energy per baryon higher than nonstrange infinite baryonic one; otherwise, protons and neutrons would decay into $u$ and $d$ quarks:
\begin{equation}
 E_{ud}/A > 930 \; \mbox{MeV}.   \label{larger930}
\end{equation}
Therefore, both, Eqs.~\ref{lower930} and ~\ref{larger930} must simultaneously satisfied. For $G_V$ = 0.3 fm$^{2}$ used in this work, the stability window lies between  139 MeV $< B^{1/4}< $ 146 MeV~\cite{Lopes_2021}. Here, we assume the maximum allowed value: $B^{1/4}$ = 146 MeV, as it will produce the lower radius for the canonical star, as well the lower value of the dimensionless tidal parameter $\Lambda$, while still producing very massive strange quark stars, $M > 2.40 \, M_\odot$.
\subsection{Superconducting CFL quark matter via  analytical approximation}
Due to the low temperature and high densities reached in the strange star interiors, the quark matter may be a color superconductor, which is a degenerate Fermi gas of quarks with a condensate of Cooper pairs near the Fermi surface that induces color Meissner effects~\cite{Alford2008}. Among the various possible configurations of superconducting matter, we can cite two possibilities: The two-flavor color-superconducting phase, where quarks with two out of three colors and two out of three flavors pair in the standard BCS fashion. The flavors with the most phase space near their Fermi surfaces, namely, $u$ and $d$, are the ones that pair, leaving the strange quark and the remaining color unpaired. Such phase is expected at densities around 2 $< n/n_0 <$ 4~\cite{Zdunik2013}. Another one is the color-flavor locked phase, where the up, down, and strange quarks can be treated on an equal footing, and the disruptive effects of the strange quark mass can be neglected. In this phase, quarks of all three colors and all three flavors form conventional spinless Cooper pairs. The CFL phase is expected at $n > 4n_0$~\cite{Zdunik2013}. For additional discussion about 2SC, CFL, and other color superconducting phases, see Ref.~\cite{Alford2008} and the references therein. 

The 2SC and the CFL phases were explored within the NJL model in Ref.~\cite{Agrawal2010}, while in Ref.~\cite{Zdunik2013}, the authors show that the color superconducting NJL EoS is very well fitted by an analytical approximation, called constant-sound-speed (CSS) parameterization, whose EoS and total number density read~\cite{Zdunik2013, Han2019, Alford2013}:
\begin{eqnarray}
p = a(\epsilon - \epsilon_{*}),  \nonumber \\ \label{SCEOS}
n = n_{*}[(1 +a)p/(a\epsilon_{*})]^{1/(1+a)} .
\end{eqnarray}

We have, therefore, three free parameters, the square of the speed of sound ($v_s^2 = a$), the energy density at $p=0$ ($\epsilon_*$), which plays a role similar to the bag in the MIT base models, and the number density at $p = 0$ ($n_*$), which in turn, plays the role of the saturation density ($n_0$) of the MIT based models. In Ref.~\cite{Zdunik2013}, the authors freely vary the value of $a$ in the range $0.2 < a < 0.8$  and found that - depending on the NJL parametrization - the 2SC phase is well described by $a < 0.33$ while the CFL phase is described by $a > 0.35$.  On the other hand, Ref.~\cite{Han2019} uses the extreme case $a = 1$. Here we consider that the quark matter is in the CFL phase and use an intermediate value, $a = 0.6$ (see the text and Fig. 4 from Ref.~\cite{Zdunik2013}, as well Ref.~\cite{Alford2013}). The value of $\epsilon_*$ is chosen as 203 MeV/fm$^3$ to match the value coming from the vector MIT bag model. Finally, $n_*$ has to be constrained, as we still need to reproduce strange quark stars in accordance with the Bodmer-Witten conjecture. We choose $n_*$ = 0.24 fm$^{-3}$, which is very close to $n_0$ = 0.23 fm$^{-3}$ coming from the vector MIT. Within this value, we have $E/A$ = 906 MeV, with implies that the analytical approximation of the  CFL satisfies Eq.~\ref{lower930} and, therefore, the Bodmer-Witten conjecture.
\section{Results and Discussions}
\subsection{Bosonic DM}
This section briefly reviews the formalism of a  bosonic DM model initially proposed in Refs.~\cite{Li_2012_1, Li_2012_2}. At very low temperatures, all particles in a dilute Bose gas condense to the same quantum ground state, forming a Bose-Einstein Condensate (BEC). Particles become correlated when their wavelengths overlap; that means the thermal wavelength is greater than the mean inter-particle distance. Assuming $T=0$ K approximation, almost all the DM particles are in the condensate. Only binary collisions at low energy are relevant in a dilute and cold gas. These collisions are characterized by a single parameter, the s-wave scattering length $l_a$, independently of the details of the two-body potential. Therefore, one can replace the interaction potential with an effective repulsive interaction:
\begin{equation}
V(\vec{r} - \vec{r'}) =  \frac{4\pi l_a}{m_x}\delta(\vec{r} - \vec{r'}),
\end{equation}
where $m_x$ is the mass of the bosonic DM.

The ground state properties of the DM are described by the mean-field Gross-Pitaevskii (GP) equation, and the equation of the state (EoS) has the form~\cite{Li_2012_1, Li_2012_2, Panotopoulos_2017, Lopes_2018}:
\begin{equation}
p_x = \frac{2\pi l_a}{m_x^3} \epsilon_x^2 . \label{bdmeos}.
\end{equation}
The scattering length $l_a$ is assumed equal to 1 fm, as in the Ref.~\cite{Li_2012_1, Li_2012_2, Panotopoulos_2017, Lopes_2018}. Moreover, the pressure strongly depends on the bosonic DM's mass due to the cubic dependence. Therefore this parameter must be taken with care. Based on the self-interaction cross-section of the DM constraint (see Refs.~\cite{Panotopoulos_2017, Lopes_2018}, the DM mass in the range 50 MeV $<m_x<$ 160 MeV. However, the original works from Ref.~\cite{Li_2012_1, Li_2012_2} suggest a mass of around 1 GeV. It is worth emphasizing that a mass ten times larger implies in pressure 1000 times lower. In Ref.~\cite{Rafiei_2022}, the authors use a slightly different model of bosonic DM, where the self-interaction is based on a scalar quartic term in the potential. They use the same constraint based on the self-interaction cross-section of the DM and suggest a mass of 400 MeV. To explore the ambiguity relative to the mass of the bosonic DM, we use here two values: 100 MeV, which agrees with Ref.~\cite{Panotopoulos_2017, Lopes_2018} and 400 MeV, which is in agreement with Ref.~\cite{Rafiei_2022}, and it is not so far from 1 GeV as suggested in Ref.~\cite{Li_2012_1, Li_2012_2}. With these settings, the pressure for $m_x$ = 400 MeV is 64 times lower than for $m_x$ = 100 MeV.

The total EoS of the strange star is, therefore, the sum of the contribution of the ordinary quark matter and the DM:
\begin{equation}
p = p_q + p_x, \quad \mbox{and} \quad \epsilon = \epsilon_q + \epsilon_x.
\end{equation}
Another important quantity is the fraction of the DM. To solve the TOV equations~\cite{Oppenheimer_1939}, we need to specify the central values both for normal matter and for DM: $p_q(0)$, $p_x(0)$ respectively. Here, we follow Ref.~\cite{Panotopoulos_2017, Lopes_2018} and define:
\begin{equation}
f_x =  \frac{p_x(0)}{p_q(0) + p_x(0)} , \label{BDMF}
\end{equation}
and use three different values for $f_x = 0.05, 0.075 $ and $0.10$. As pointed out in Ref.~\cite{Panotopoulos_2017, Lopes_2018}, these values agree with the current DM constraints obtained from stars like the Sun.
\subsection*{1. Bosonic DM within vector MIT bag model}
\begin{figure}
\centering
\includegraphics[width=0.32\textwidth,angle=270]{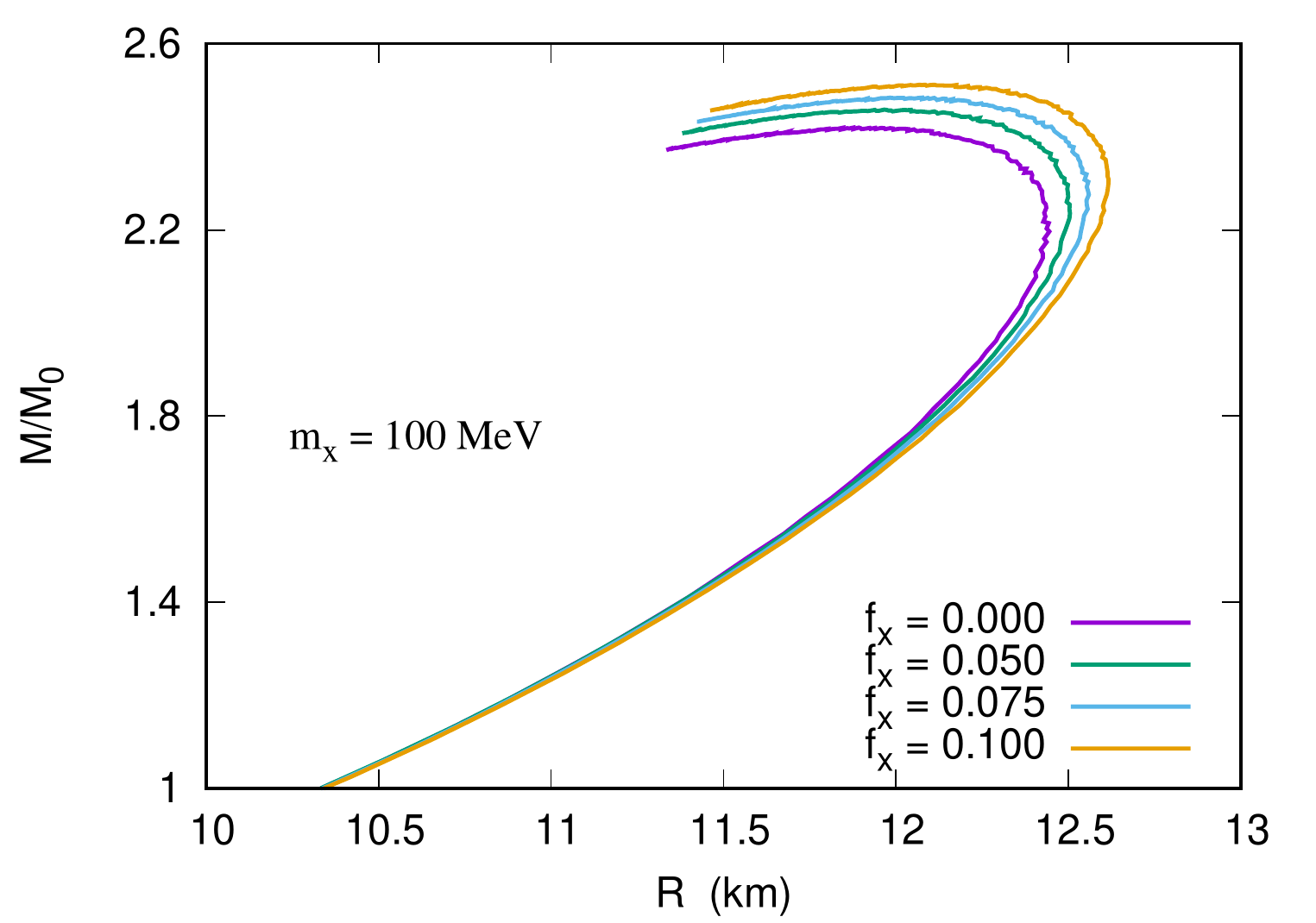}
\includegraphics[width=0.32\textwidth,,angle=270]{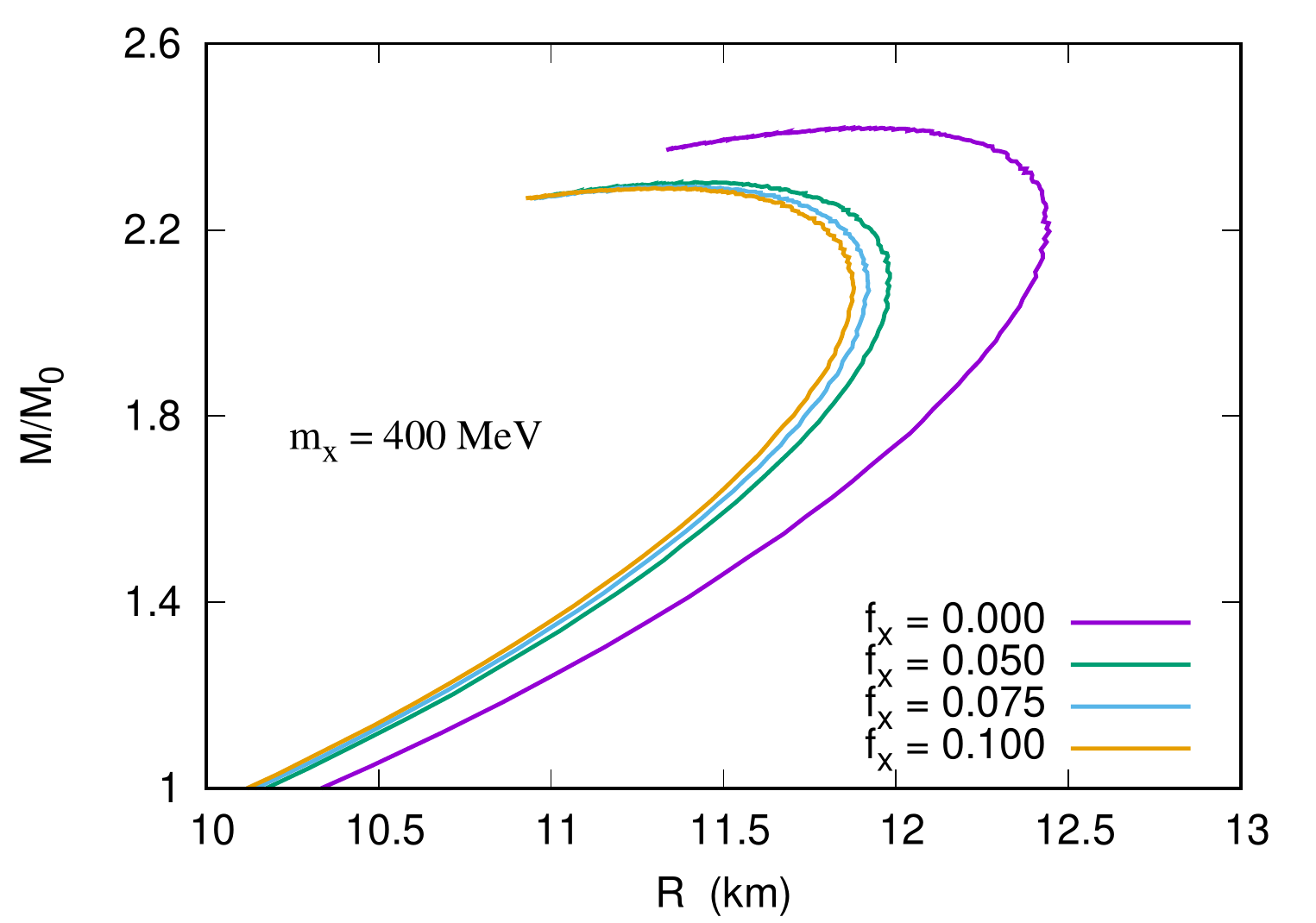}
\caption{Mass-radius relation for bosonic DM admixed strange stars with $m_x$ =100 MeV (left) and $m_x$ = 400 MeV (right).} \label{F1}
\end{figure}

In Fig.~\ref{F1}, we plot the TOV solution for bosonic DM admixed strange stars with the mass of 100 MeV and 400 MeV.

As can be seen, for a bosonic DM mass of 100 MeV, we have an increase in the maximum mass with the increase of the fraction of DM. This result is coherent with those presented in Ref.~\cite{Panotopoulos_2017, Lopes_2018} for the original, massless MIT. Moreover, as in the case of the original massless MIT, with the massive vector MIT, we also see that the presence of DM affects only massive stars. Strange stars with $M < 1.5 \, M_\odot$ reproduced essentially the same radii. The maximum masses vary from 2.41 $M_\odot$ for pure strange stars to 2.51$M_\odot$ for bosonic DM admixed with a fraction of 0.10. This indicates that the PSR J0740+6620 with a gravitational mass of 2.08 $\pm$ 0.07 $M_\odot$~\cite{Fonseca_2021} can indeed be a stable strange star with or without admixed bosonic DM. Even the possible mass of 2.35 $\pm$ 0.17 $M_\odot$ of the black widow pulsar PSR J0952-0607~\cite{Romani_2022} can be explained as bosonic DM matter admixed strange star. On the other hand, the radius of the canonical star is in the narrow range of 11.37 km to 11.40 km. In the literature, there is no consensus about the true value of the radius of the canonical star. For instance, in ref.~\cite{Ozel_2016}, the constraint on the radius of the canonical star is $10.1-11.1$ km, which provides a very narrow range. If this is true, neither of our results can fulfill such tight constraints. In Ref.~\cite{Capano_2020}, an upper limit of 11.9 km was provided. In this case, our results are in full agreement. However, recent results from the NICER x-ray telescope point that the radius of the canonical star is between 11.52 km and 13.85 km ~\cite{Riley_2019} and between 11.96 km and 14.26 km as given in Ref.~\cite{Miller_2019}. In these cases, our radii are too small.

Now, we have opposite results for a  mass $m_x$ = 400 MeV. First, the maximum mass decrease with the increase of DM fraction, dropping from $2.41 \, M_\odot$ to $2.29 \, M_\odot$ for a fraction $f_x$ of 0.10. However, all values agree with the mass of the PSR J0740+6620~\cite{Fonseca_2021} and the PSR J0952-0607~\cite{Romani_2022}. Secondly, we see that even low-mass strange stars are already affected by the DM and are significantly more compact. The radius of the $1.4 \, M_\odot$ strange star can reach a value as low as 11.08 km. Therefore, this result is in agreement with both Refs.~\cite{Ozel_2016, Capano_2020}. The polytropic EoS of Eq.~\ref{bdmeos} can easily explain these results. A four times higher DM matter mass produces sixty-four times smaller pressure. The reduction of the pressure causes the reduction of the maximum mass and increases the star compression. 

\begin{figure}
\centering
\includegraphics[width=0.32\textwidth,angle=270]{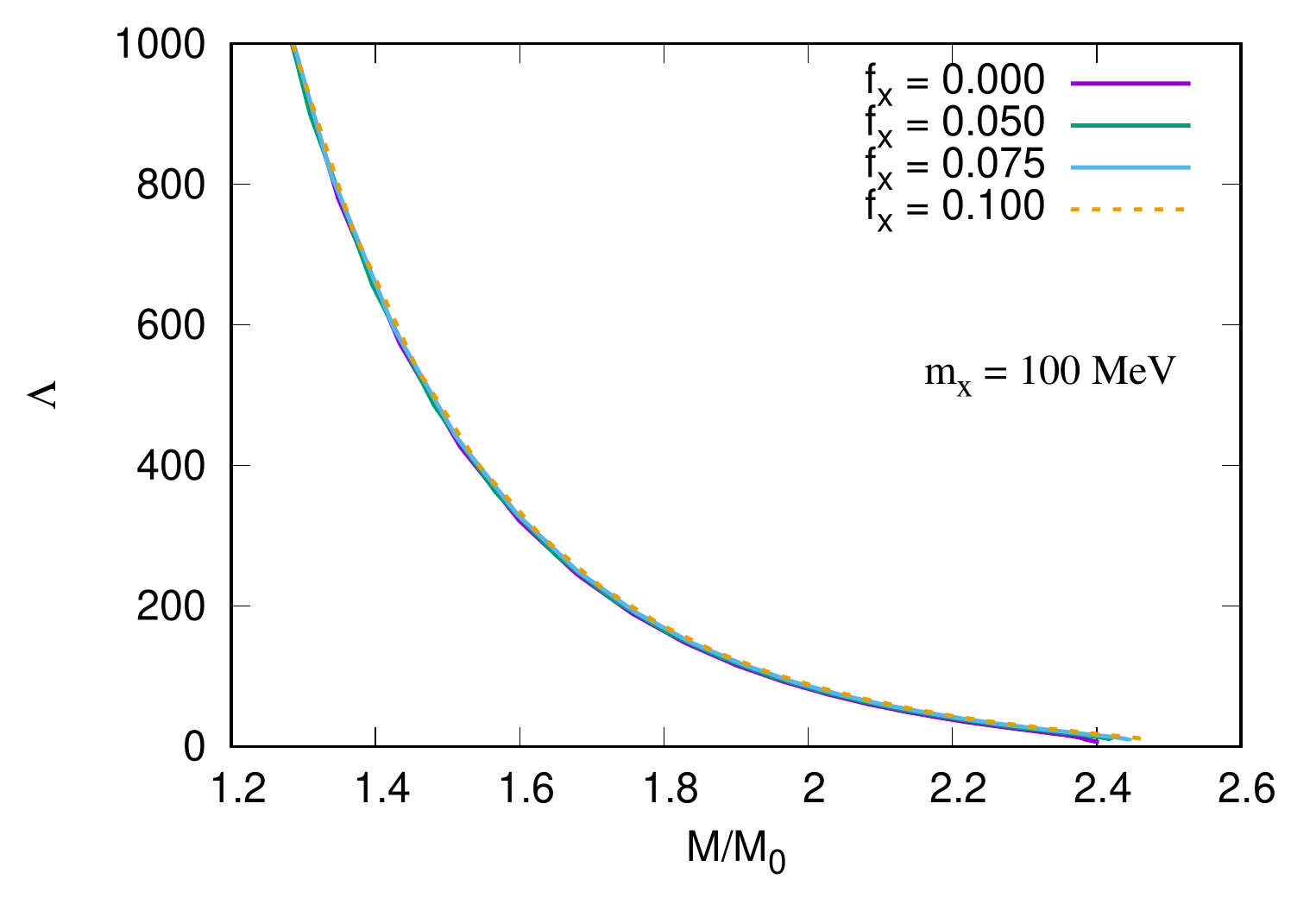}
\includegraphics[width=0.32\textwidth,,angle=270]{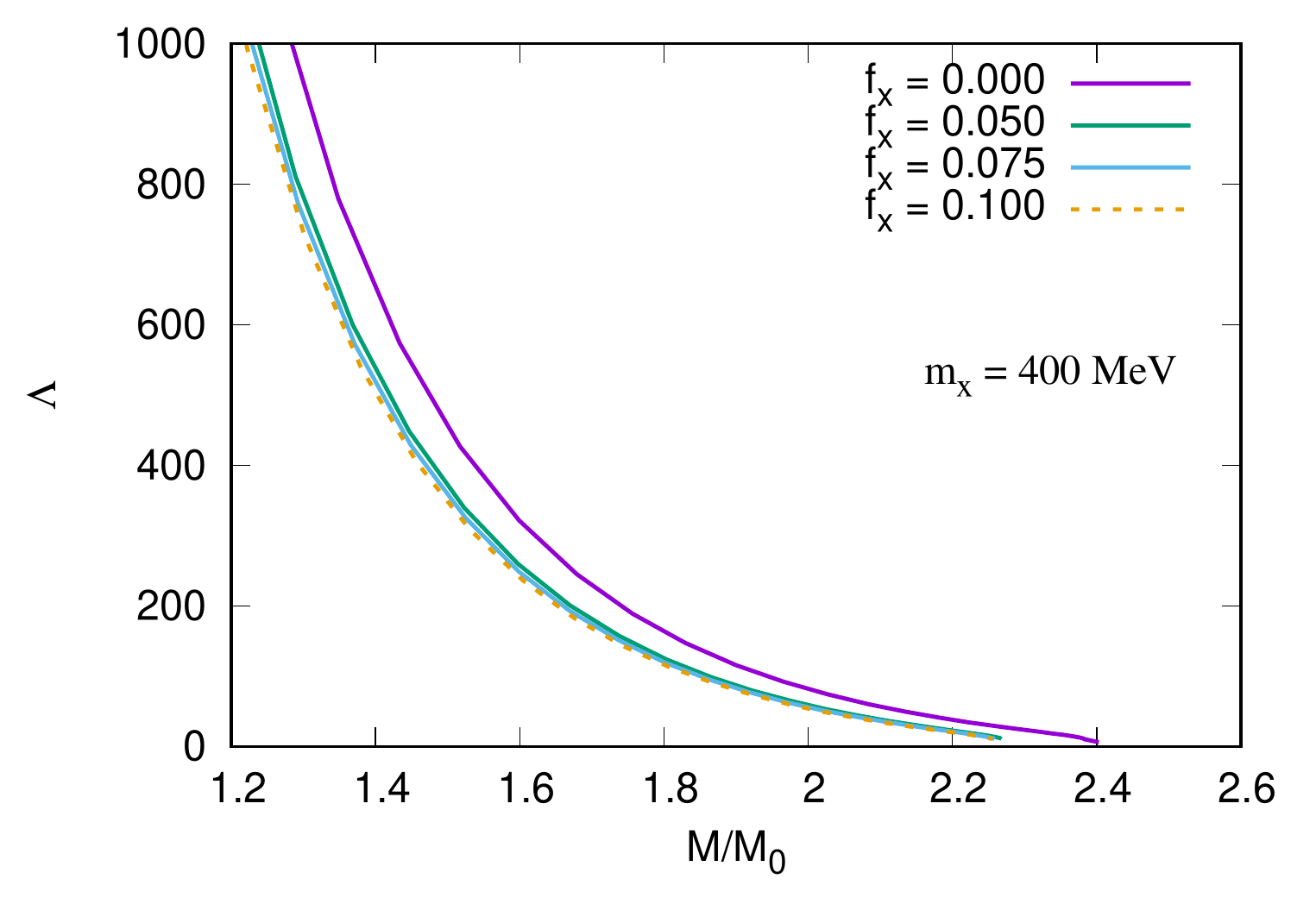}
\caption{Dimensionless tidal parameter $\Lambda$ for bosonic DM admixed strange stars with $m_x=100$ MeV (left) and $m_x=400$ MeV (right).}
\label{F2}
\end{figure}

Another essential quantity and constraint is the so-called dimensionless tidal deformability parameter $\Lambda$. If we put an extended body in an inhomogeneous external field, it will experience different forces throughout its surface. The result is a tidal interaction. The tidal deformability of a compact object is a single parameter $\lambda$ that quantifies how easily the object is deformed when subjected to an external tidal field. Larger tidal deformability indicates that the object is easily deformed. Conversely, a compact object with a small tidal deformability parameter is more compact and more difficult to deform. The tidal deformability is defined as:
\begin{equation}
 \Lambda~\equiv~\frac{\lambda}{M^5}~\equiv~\frac{2k_2}{3C^5} , \label{tidal}
\end{equation}
where $M$ is the compact object mass and $C = GM/R$ is its compactness. The parameter $k_2$ is called the second (order) Love number (see the appendix).  Nevertheless, as pointed out in Refs.~\cite{Postnikov_2010, Lourenco_2021}, the value of $y_R$ must be corrected since strange stars are self-bound and present a discontinuity at the surface. Therefore we must have
\begin{equation}
 y_R \rightarrow y_R - \frac{4\pi R^3 \Delta\epsilon_S}{M} , \label{yr}
\end{equation}
where $R$ and $M$ are the star radius and mass, respectively, and $\Delta\epsilon_S$ is the difference between the energy density at the surface ($p=0$) and the star's exterior (which implies $\epsilon=0$). The results for the dimensionless tidal parameter are displayed in Fig.~\ref{F2}.

As we can be seen, some features present in the mass-radius relation are also present here. For instance, for a mass $m_x$ = 100 MeV, the low masses of strange stars have similar tidal parameters, despite their DM fraction. The tidal parameter for the canonical mass lies between 638 and 644. These values are in agreement with the constraint $\Lambda <$ 800~\cite{Abbott_2017}, but fail to fulfill the constraint 
70 $< \Lambda <$ 580~\cite{Abbott_2018}.

In the case of $m_x$ = 400 MeV, the strange stars' huge compression due to an increase in the DM fraction reduces the tidal parameter $\Lambda$. The tidal parameter now lies around 500. This indicates that for $m_x$ = 400 MeV, we are able to explain very massive neutron stars as the PSR J0952-0607~\cite{Romani_2022}, and simultaneously fulfills the constraints of $\Lambda <$ 800~\cite{Abbott_2017} and 70 $< \Lambda <$ 580~\cite{Abbott_2018}.  We summarize the results of this section in Tab~\ref{T1}.
\begin{table}
\centering
\caption{Macroscopic properties of bosonic DM admixed strange stars }
\begin{tabular}{ccccccc}
\hline \hline
$m_x$ (MeV)  &  $f_x$ & $M/M_\odot$ & R (km) & $R_{1.4}$ (km) & $\Lambda_{1.4}$ \\ \hline
100  & 0.000  & 2.41 & 11.86 & 11.37 &  644 \\ 
100  & 0.050  & 2.46 & 12.01 & 11.37 &  638 \\ 
100  & 0.075  & 2.48 & 12.06 &11.38 &  645  \\
100  & 0.100  & 2.51 & 12.08 & 11.40 & 652  \\
\hline
400 &  0.000  & 2.41 & 11.86 & 11.37 & 644  \\
400  & 0.050   &2.31 & 11.42 & 11.16 & 526  \\
400  & 0.075  & 2.30 & 11.38 & 11.12 & 497  \\
400  & 0.100  & 2.29 & 11.31 & 11.08 & 480  \\
\hline \hline 
\end{tabular}
\label{T1}
\end{table}
\subsection*{2. Bosonic DM within CFL quark matter}
\begin{figure}
\centering
\includegraphics[width=0.32\textwidth,angle=270]{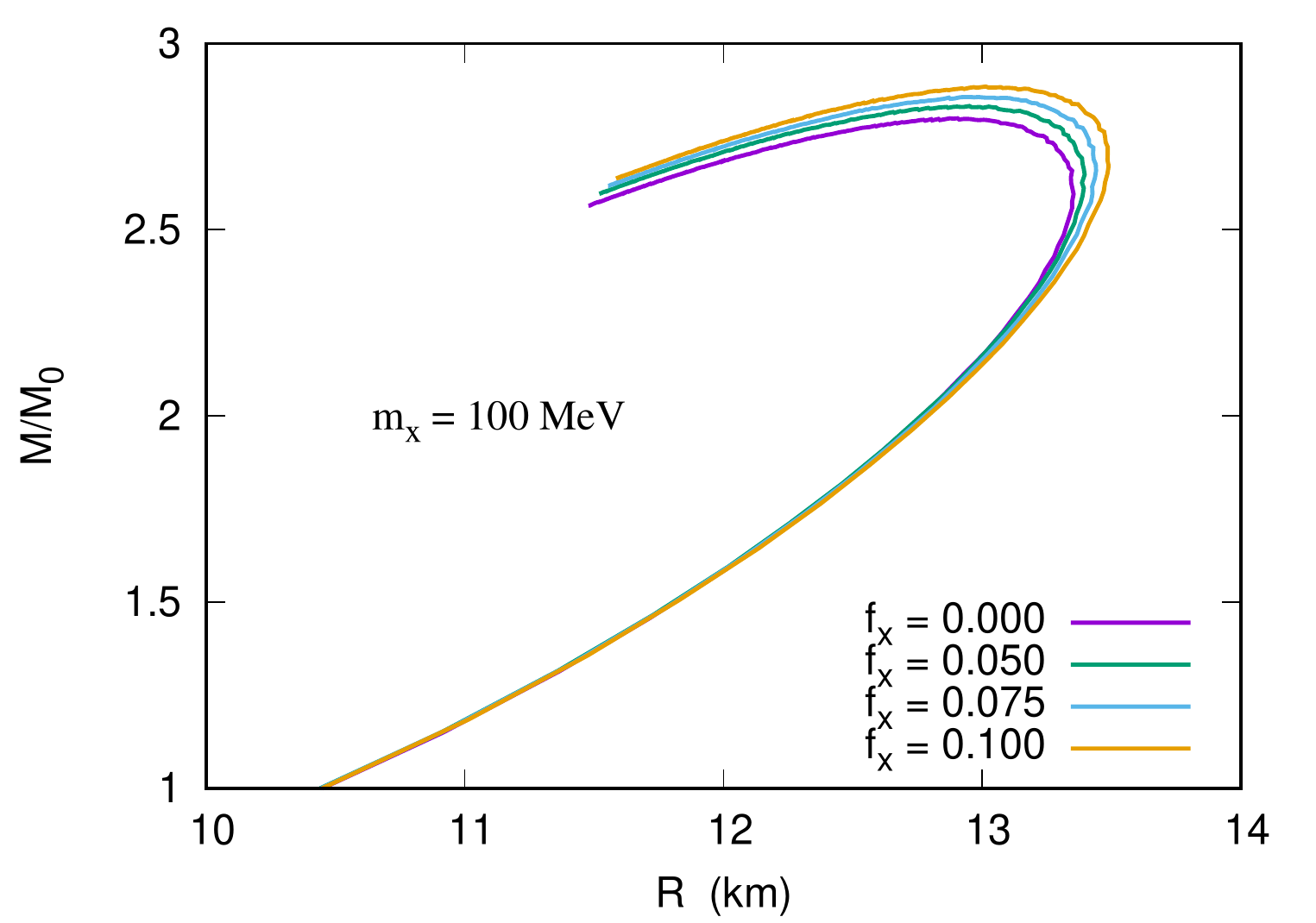}
\includegraphics[width=0.32\textwidth,,angle=270]{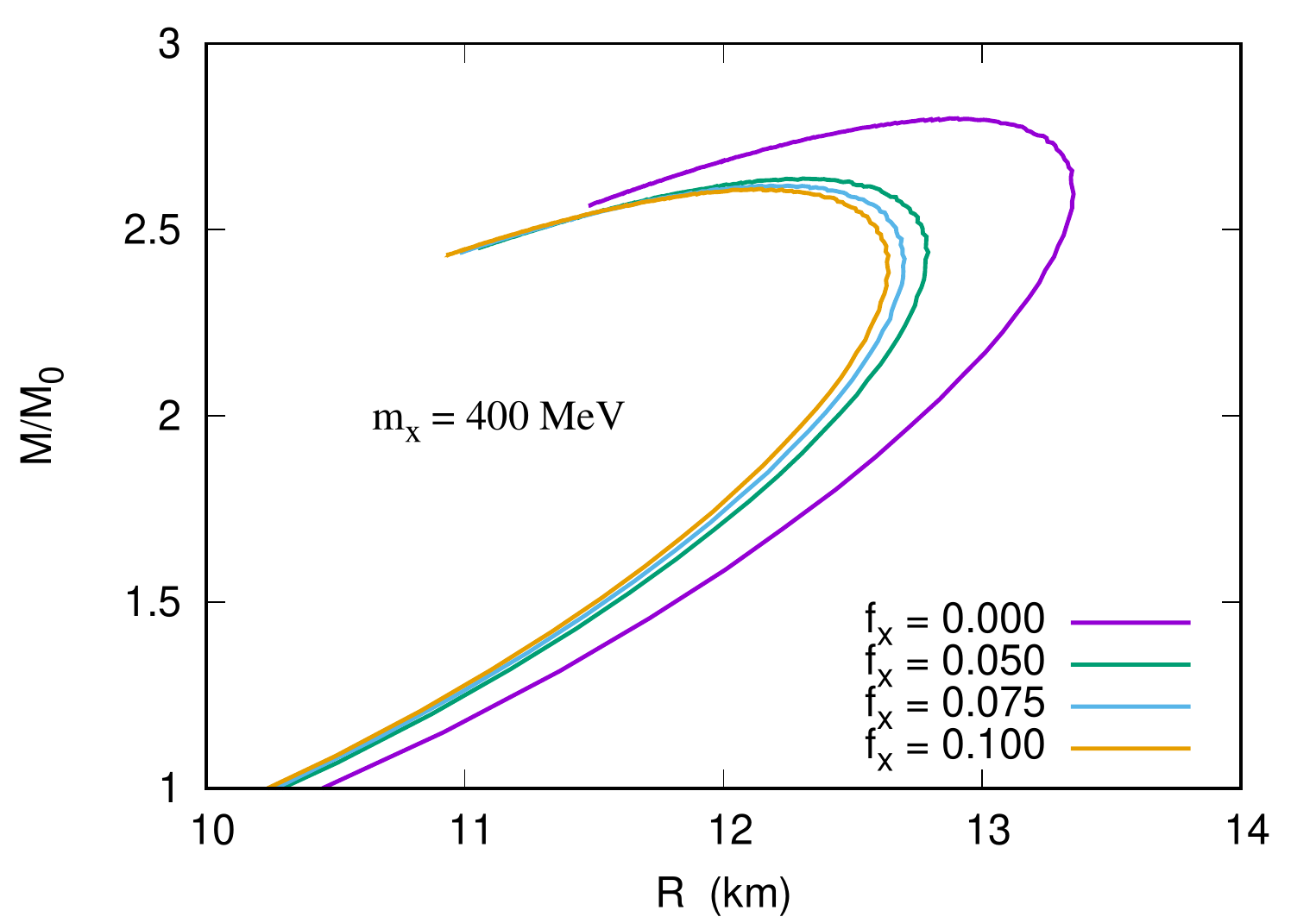}
\caption{Mass-radius relation for bosonic DM admixed  CFL strange stars with $m_x$ =100 MeV (left) and $m_x$ = 400 MeV (right).} \label{FSC1}
\end{figure}

In order to better understand the effects of the DM in strange stars, we now assume that the quark matters are in the CFL superconducting phase via the analytical approximation EoS in Eq.~\ref{SCEOS}. The mass-radius relations are presented in Fig.~\ref{FSC1}.

As can be seen, for a bosonic DM mass of 100 MeV, we have an increase in the maximum mass with the increase of the fraction of DM. The qualitative results for CFL superconducting quark stars are analogous to both the original, massless MIT as showed in Ref.~\cite{Panotopoulos_2017, Lopes_2018}, as well for the massive vector MIT bag model as presented in the last section. This indicates a possible model-independent behavior about the effect of the bosonic DM. Moreover, as in the case of the original massless MIT and the massive vector MIT,  in the CFL phase, we also see that the presence of DM affects only massive stars. CFL strange stars with $M < 1.8 \, M_\odot$ reproduced essentially the same radii. The maximum masses vary from 2.81 $M_\odot$ for pure strange stars to 2.88$M_\odot$ for bosonic DM admixed with a fraction of 0.10. This indicates that the PSR J0740+6620 with $M$ = 2.08 $\pm$ 0.07 $M_\odot$~\cite{Fonseca_2021} can be a stable CFL strange star with or without admixed bosonic DM. Even the possible mass of 2.35 $\pm$ 0.17 $M_\odot$ of the pulsar PSR J0952-0607~\cite{Romani_2022} can be explained as bosonic DM matter admixed strange star. On the other hand, the radius of the canonical star presents almost no variation and is fixed at around 11.57 km. Such a value is too low to reproduce the constraint range of   $10.1-11.1$ km, shown in Ref.~\cite{Ozel_2016} while agreeing with Ref.~\cite{Capano_2020}, whose upper limit is 11.9 km. About the  NICER x-ray telescope, the constraint between 11.52 km and 13.85 km pointed in Ref.~\cite{Riley_2019} is fulfilled, but the bound in the range between 11.96 km and 14.26 km  (Ref.~\cite{Miller_2019}) is not. 

For a  mass $m_x$ = 400 MeV, the results for CLF superconducting strange stars are analogous to the massive MIT bag model discussed in the last section. The maximum mass decrease with the increase of DM fraction, dropping from $2.81 \, M_\odot$ to $2.61 \, M_\odot$ for a fraction $f_x$ of 0.10. However, all values agree with the mass of the PSR J0740+6620~\cite{Fonseca_2021} and the black widow pulsar PSR J0952-0607~\cite{Romani_2022}. Secondly, we see that even low-mass strange stars are already affected by the DM and are significantly more compact. The radius of the $1.4 \, M_\odot$ for $f_x = $  0.10 is about  11.29 km. Such a low radius fails to fulfill both NICER constraints~\cite{Riley_2019,Miller_2019}, but is in agreement with Capano {\it et al.} ~\cite{Capano_2020}. The reduction of the CFL strange star and its compression can again be explained by the polytropic EoS of Eq.~\ref{bdmeos}. A four times higher DM matter mass produces sixty-four times smaller pressure. The reduction of the pressure causes the reduction of the maximum mass and increases the star compression. 
\begin{figure}
\centering
\includegraphics[width=0.32\textwidth,angle=270]{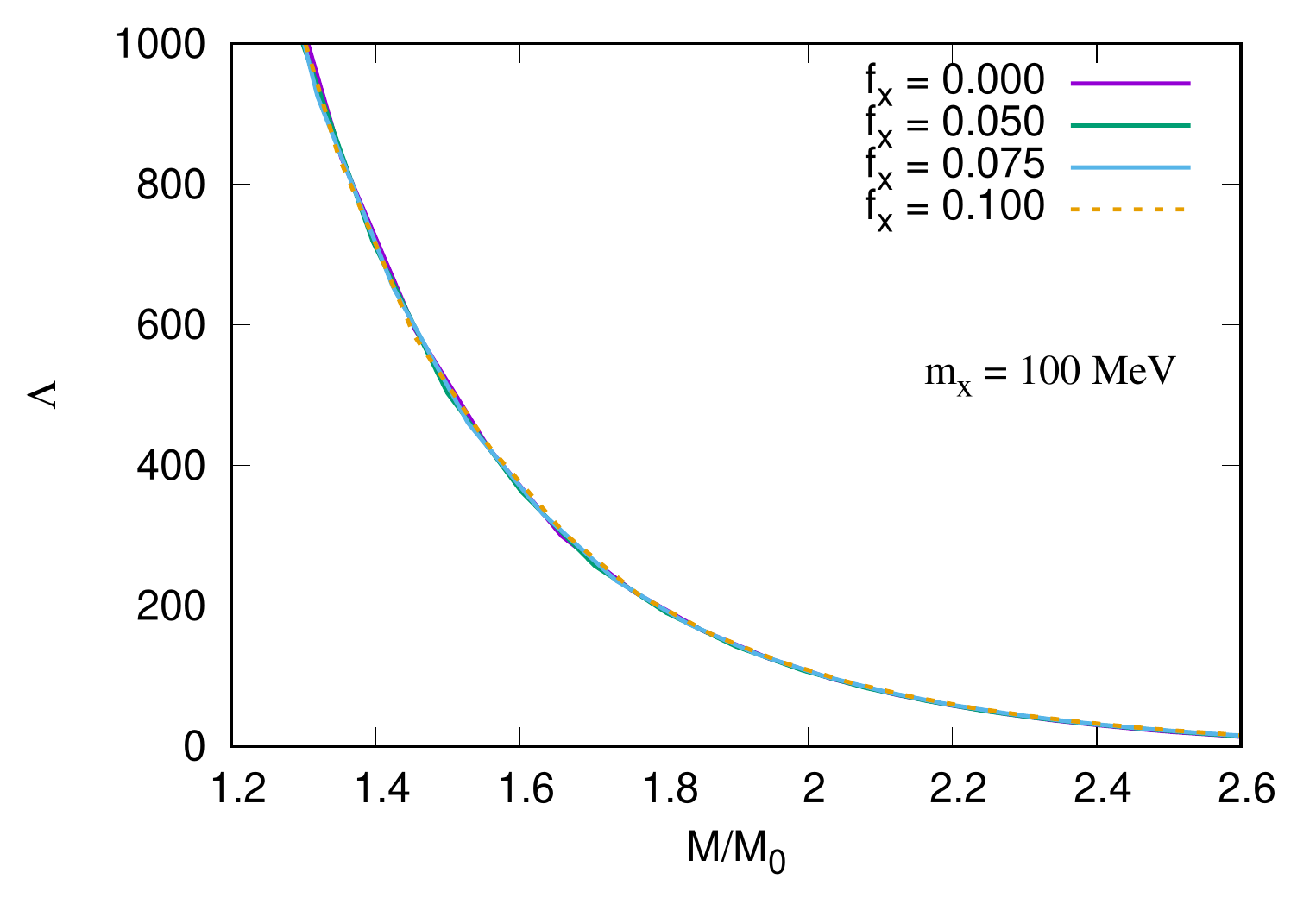}
\includegraphics[width=0.32\textwidth,,angle=270]{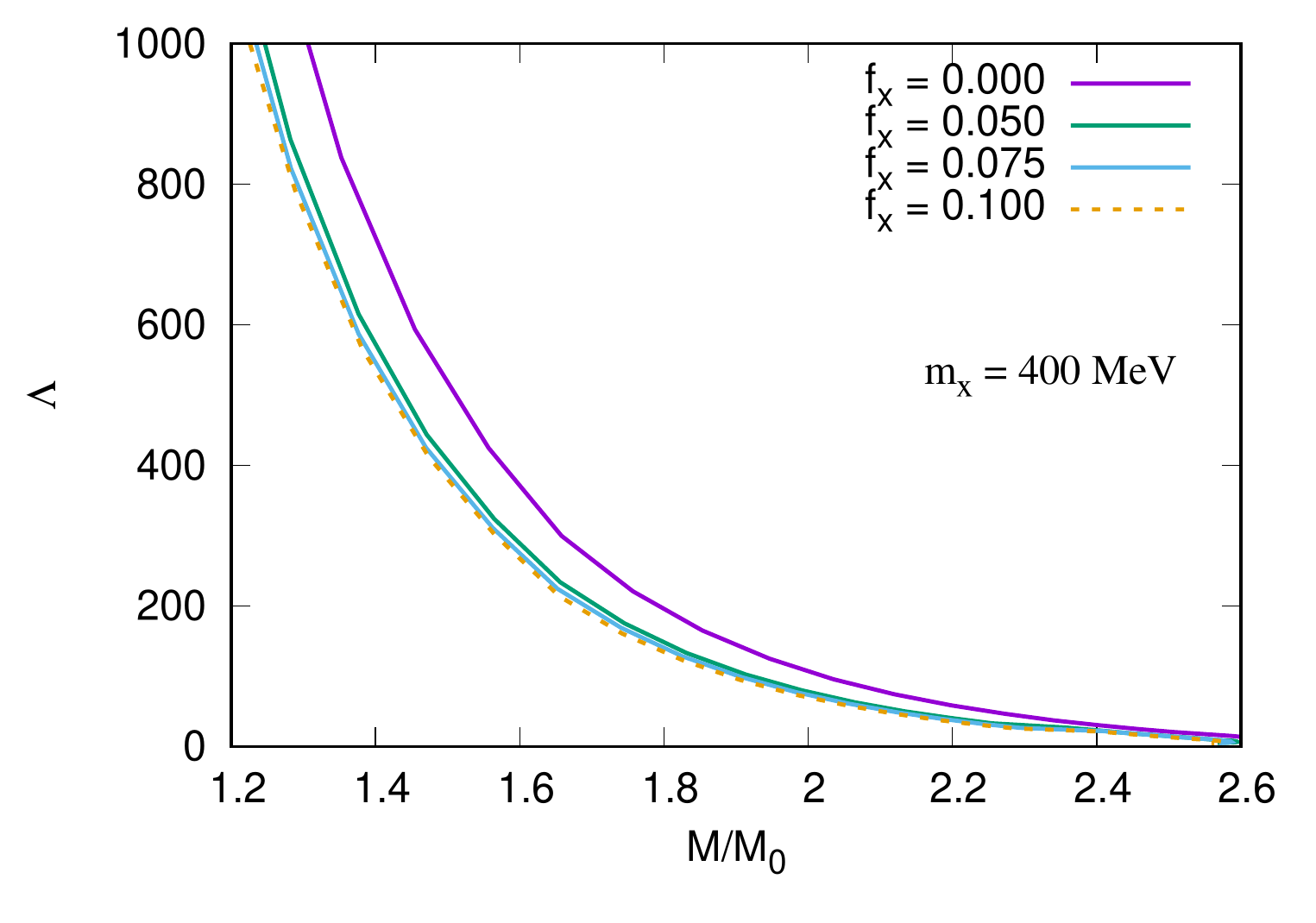}
\caption{Dimensionless tidal parameter $\Lambda$ for bosonic DM admixed CFL superconducting strange stars with $m_x=100$ MeV (left) and $m_x=400$ MeV (right).}
\label{FSC2}
\end{figure}

We also calculate the dimensionless tidal parameter $\Lambda$ for the CFL superconducting strange stars. The results are presented in Fig.~\ref{FSC2}.
As we can be seen, the results are analogous to the vector MIT bag model. As in the case of the mass-radius relation, for low mass stars there is very low variation in the $\Lambda$. For instance, for a mass $m_x$ = 100 MeV, the low masses strange stars have similar tidal parameters, despite their DM fraction. The tidal parameter for the canonical mass lies between 709 and 721. These values are in agreement with the constraint $\Lambda <$ 800~\cite{Abbott_2017}, but fail to fulfill the constraint 70 $< \Lambda <$ 580~\cite{Abbott_2018}.

In the case of $m_x$ = 400 MeV, the results for CFL superconducting strange stars are again analogous to vector MIT strange stars. The stars' huge compression as the DM fraction increases reduce the tidal parameter $\Lambda$. The tidal parameter now lies around 550. This indicates that for $m_x$ = 400 MeV, we are able to explain very massive neutron stars as the PSR J0952-0607~\cite{Romani_2022}, and simultaneously fulfills the constraints of $\Lambda <$ 800~\cite{Abbott_2017} and 70 $< \Lambda <$ 580~\cite{Abbott_2018}.  We summarize the results of this section in Tab~\ref{TSC1}.
\begin{table}
\centering
\caption{Macroscopic properties of bosonic DM admixed CFL superconducting strange stars}
\begin{tabular}{ccccccc}
\hline \hline
$m_x$ (MeV)  &  $f_x$ & $M/M_\odot$ & R (km) & $R_{1.4}$ (km) & $\Lambda_{1.4}$ \\ \hline
100  & 0.000  & 2.81 & 12.89 & 11.57 &  721 \\ 
100  & 0.050  & 2.83 & 12.84 & 11.57 &  709 \\ 
100  & 0.075  & 2.86 & 12.96 &11.58 &  717  \\
100  & 0.100  & 2.88 & 13.00 & 11.58 & 717  \\
\hline
400 &  0.000  & 2.81 & 12.89 & 11.57 & 721  \\
400  & 0.050   &2.63 & 12.30 & 11.37 & 570  \\
400  & 0.075  & 2.62 & 12.22 & 11.32 & 545  \\
400  & 0.100  & 2.61 & 12.13 & 11.29 & 531  \\
\hline \hline 
\end{tabular}
\label{TSC1}
\end{table}
\subsection{Fermionic DM}
The Lagrangian of the fermionic DM reads~\cite{Das_2021,OdilonDM,Lourenco_2021}:
\begin{eqnarray}
\mathcal{L}_{\rm DM} &=& \bar{\chi}(i \gamma^\mu \partial_\mu - (m_x -g_H h))\chi 
+ \frac{1}{2}(\partial^\mu h \partial_\mu h - m_H^2 h^2). \label{FDMEOS}
\end{eqnarray}
Here, we assume a dark fermion represented by the Dirac field $\chi$ that self-interacts through the exchange of the Higgs boson, whose mass is $m_H$ = 125 GeV. The coupling constant is assumed to be $g_H = 0.1$,  which agrees with the constraints in Refs.~\cite{Panotopoulos_2017, Das_2021}. Within this prescription, the DM self-interaction is very feeble and behaves as a free Fermi gas. More explicitly, the strength of the interaction is:
\begin{equation}
G_H = \bigg ( \frac{g_H}{m_H} \bigg )^2 = 2.492 \times 10^{-8} \quad \mbox{fm}^{2} . \label{higgsc}
\end{equation}
\begin{figure}
\centering
\includegraphics[width=0.32\textwidth,angle=270]{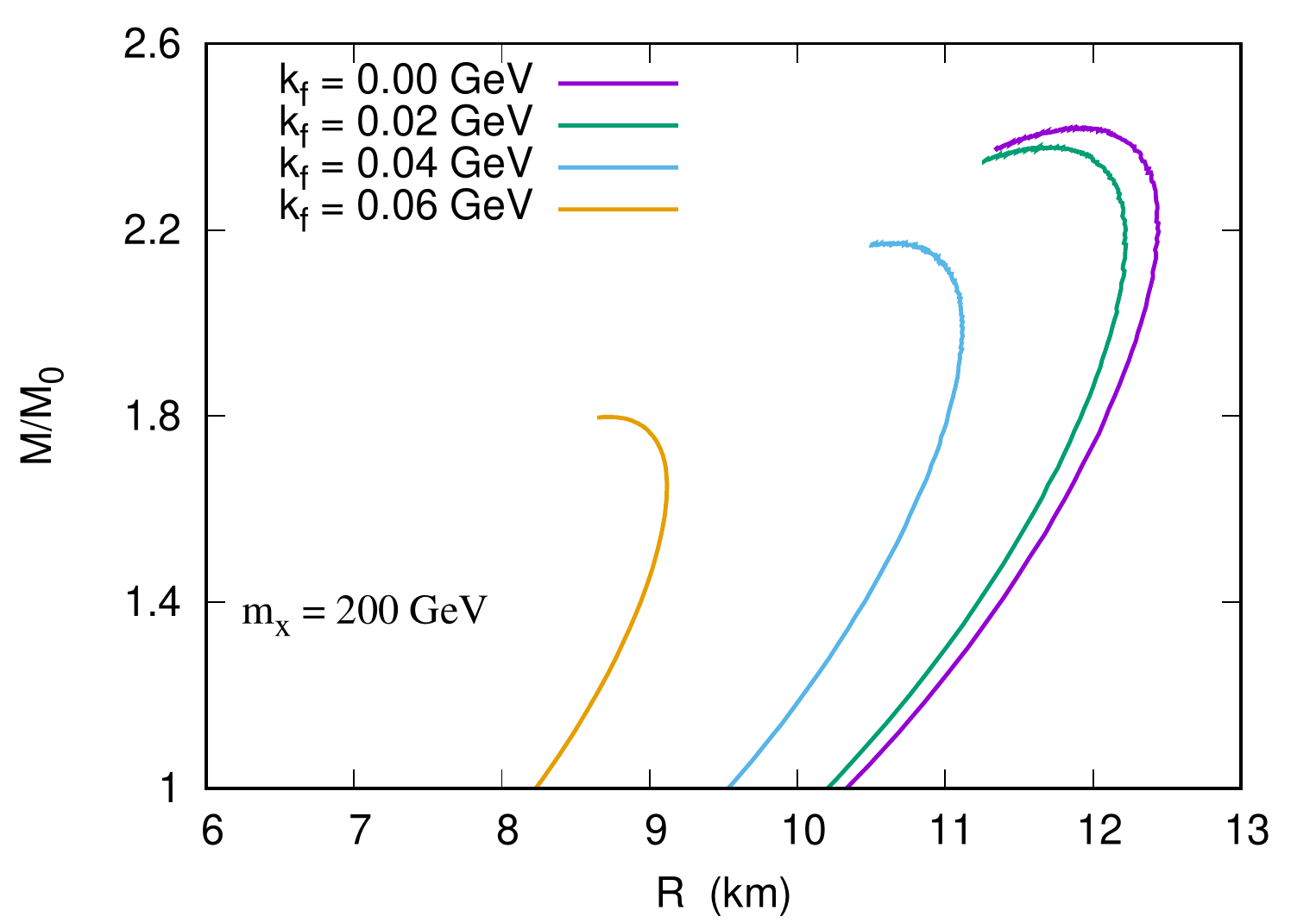}
\includegraphics[width=0.32\textwidth,,angle=270]{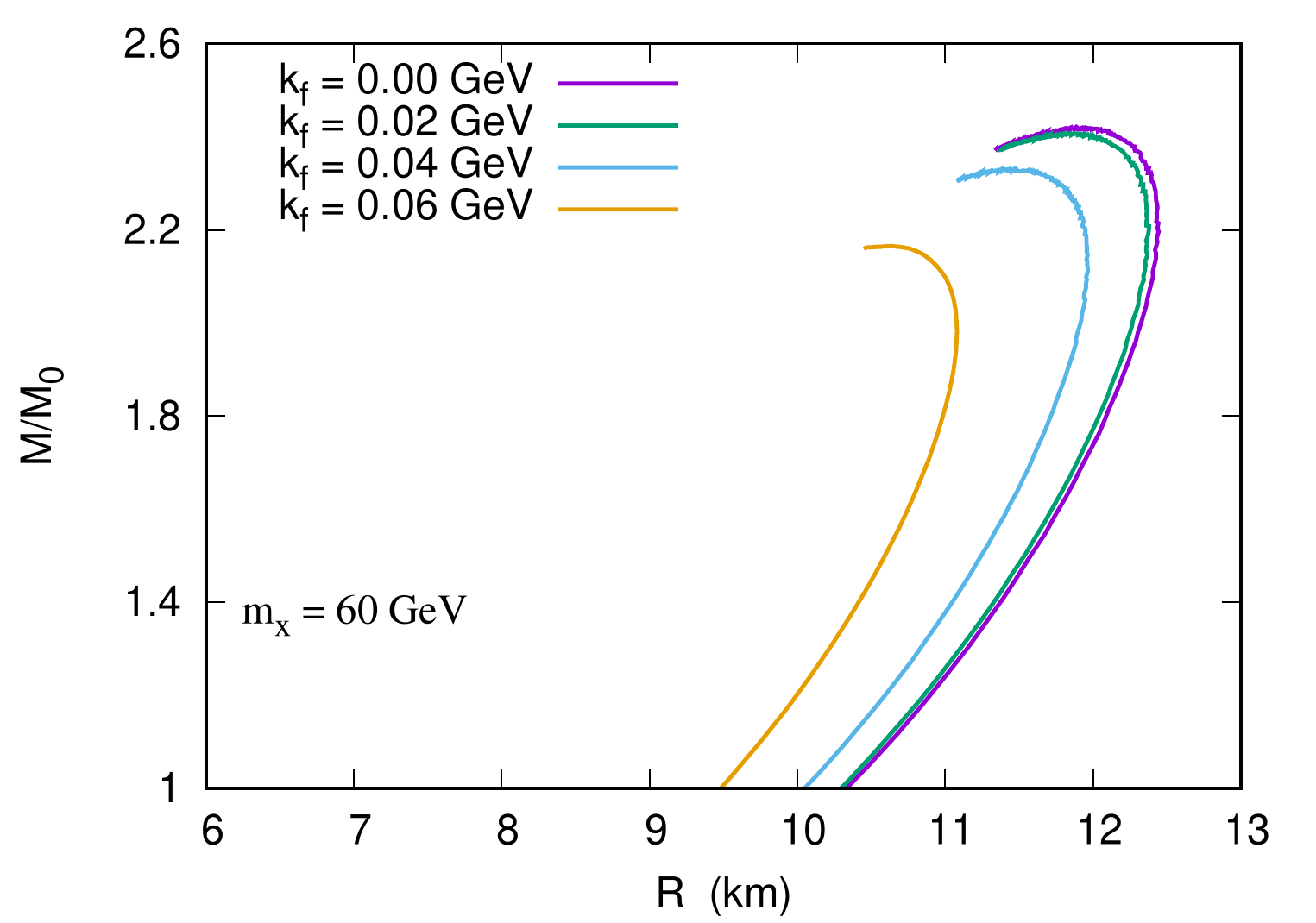}
\caption{Mass-radius relation for fermionic DM admixed strange stars with $m_x =200$ GeV (left) and $m_x = 60$ GeV (right).}
\label{F3}
\end{figure}

The EoS is easily obtained in mean field approximation, completely analogous to the QHD model~\cite{Serot_1992}. The fermionic DM is assumed to be the lightest neutralino, with $m_x$ = 200 GeV, as done in Ref.~\cite{OdilonDM, Das_2021}. However, as pointed out in Ref.~\cite{Lopes_2002}, the lower limit for weakly interacting massive particles (WIMP) is 60 GeV. Therefore we also use $m_x$ = 60 GeV to better study the influence of the DM mass. As in the case of the bosonic DM, we must fix the DM fraction. As we are dealing here with fermionic DM, we follow ref.~\cite{Das_2021,Guha_2021,OdilonDM} and use the Fermi momentum to fix the DM fraction, using three different values: $k_f^{\rm DM}$ = 0.02 GeV, 0.04 GeV, and 0.06 GeV. 
\subsection*{1. Fermionic DM within vector MIT bag model}
We display in Fig.~\ref{F3} the TOV solution for a fermionic DM with a mass of 200 GeV and 60 GeV within the vector MIT bag model. As can be seen, the results for fermionic DM are significantly different when compared with bosonic DM. The maximum masses are always reduced, and the star compression always increases, even for very low masses. Also, different DM fractions always produce different mass-radius relations, affecting all the strange star families, unlike the bosonic case, where we have very similar stars for different DM fractions, which is easily understood by the different criteria of the DM fraction. In the case of bosonic DM, the DM fraction is dependent on the quark EoS via Eq.~\ref{BDMF}. In the case of fermionic DM, the Fermi momentum is fixed and independent of the quark EoS.

Qualitatively, the results for $m_x$ = 200 GeV and 60 GeV are the same. Increasing the DM fraction compress the star and reduces the maximum mass. Quantitatively, we see that a higher DM mass has a strong influence once it has a higher increase in the energy density, and at the same time, that produces a lower contribution to the pressure. The maximum mass drops from 2.41 $M_\odot$ for $k_f^{\rm DM}$ = 0.00 to only 1.80 $M_\odot$ for $k_f^{\rm DM}$ = 0.06 GeV in the case of $m_x$ = 200 GeV and to 2.16 $M_\odot$ for $m_x$ = 60 GeV. In the same sense, the radius of the canonical star drops from 11.37 km for $k_f^{\rm DM}$ = 0.00 to only 8.95 km for $k_f^{\rm DM}$ = 0.06 GeV in the case of $m_x$ = 200 GeV, and to 10.42 km for $m_x$ = 60 GeV. As can be seen, the results for $k_f^{\rm DM}$ = 0.06 GeV with $m_x$ = 200 GeV can be ruled out once it has a very low maximum mass in disagreement with the NICER result of the PSR J0740+6620 with a gravitational mass of 2.08 $\pm$ 0.07 $M_\odot$~\cite{Fonseca_2021}, and also a very low radius for the canonical star, in disagreement even with the low limit of 10.1 km presented in Ref.~\cite{Ozel_2016}.

\begin{figure}
\centering
\includegraphics[width=0.333\textwidth,angle=270]{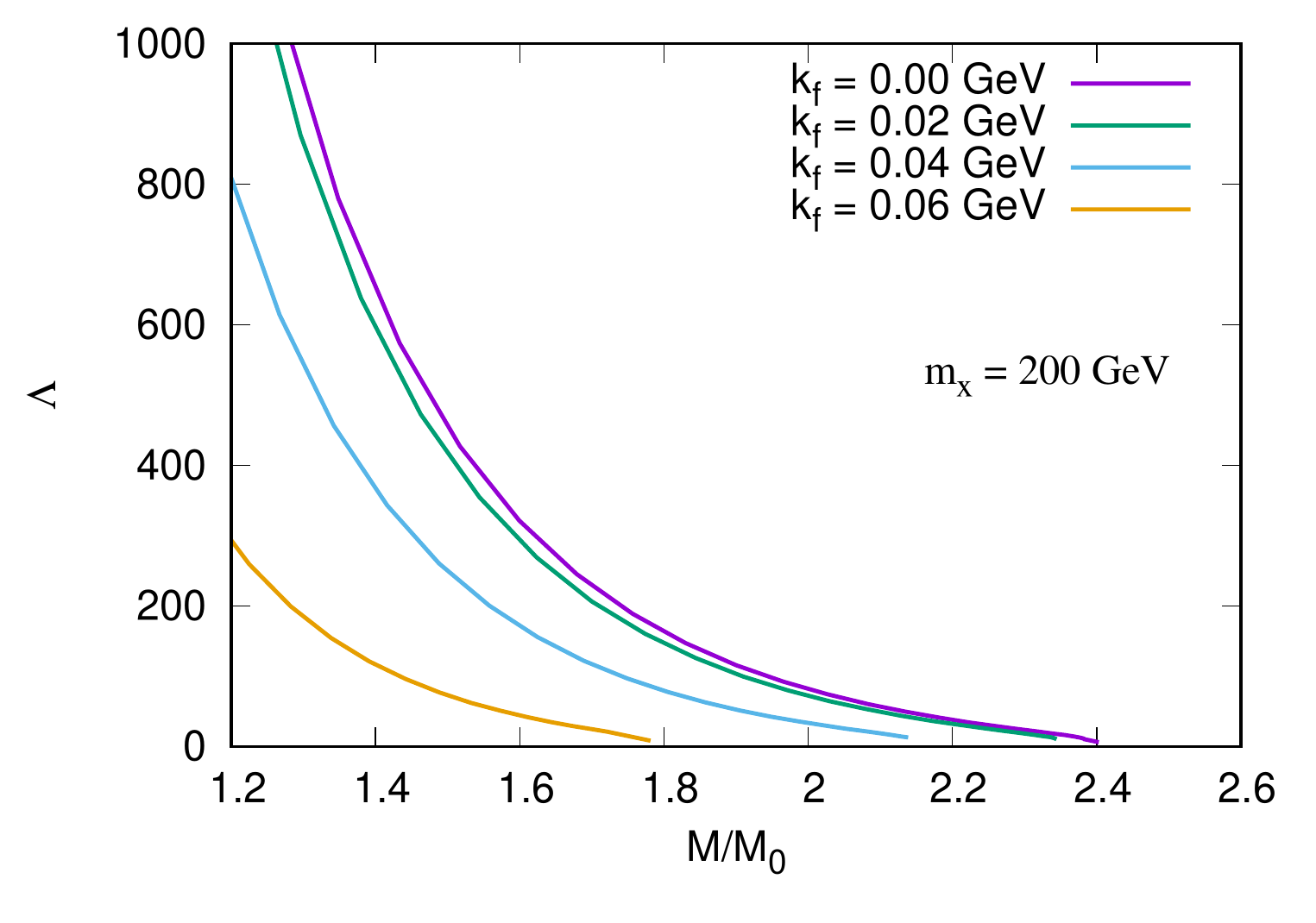}
\includegraphics[width=0.333\textwidth,angle=270]{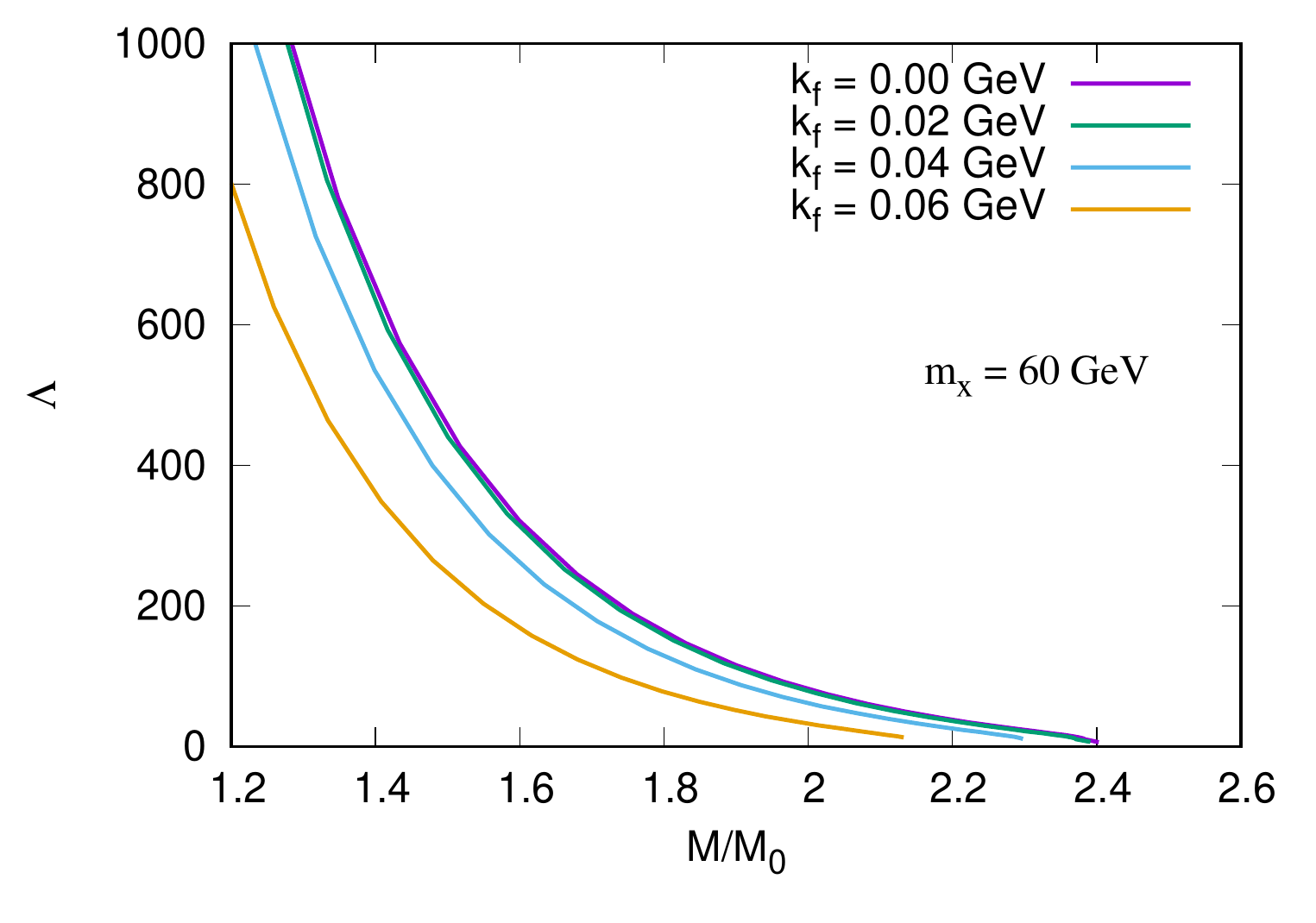}
\caption{Dimensionless tidal parameter $\Lambda$ for fermionic DM admixed strange stars with $m_x=200$ GeV (left) and $m_x = 60$ GeV (right).} 
\label{F4}
\end{figure}

We plot in Fig.~\ref{F4} the dimensionless parameter $\Lambda$ for fermionic DM admixed strange stars with $m_x$ = 200 GeV and $m_x$ = 60 GeV within the vector MIT bag model. As we can see, the strong compression due to the fermionic DM contribution reduces the tidal parameter significantly. In the case with $m_x$ = 200 GeV and $k_f^{\rm DM}$ = 0.06 GeV, the tidal parameter drops to only 108, which is six times lower than for $k_f^{\rm DM}$ = 0.00, although, as we pointed out before, such parametrization must be ruled out. 

As can be seen, most of the parametrizations are able to fulfill the main constraints for pulsar observations, i.e., $M > 2.01 M_\odot$ and $70 ~<~\Lambda~< 580$. Indeed, the presence of DM improves the theoretical prediction and the observational constraints, although it can be some debate about the radius of the canonical star. They do not fulfill NICER results~\cite{Miller_2019,Riley_2019} but agree with Ref.~\cite{Capano_2020}.

It is also worth to point the existence of almost degenerate results. As can be seen, for $m_x$ = 200 GeV with $k_f^{\rm DM}$ = 0.04 GeV, the macroscopic are essentially the same for the  $m_x$ = 60 GeV and
$k_f^{\rm DM}$ = 0.06 GeV. The main results are summarized in Tab.~\ref{T2}.
\begin{table}
\centering
\caption{Macroscopic properties of fermionic DM admixed  strange stars}
\begin{tabular}{ccccccc}
\hline \hline
$m_x$ (GeV)  &  $k_f^{\rm DM}$ (GeV) & $M/M_\odot$ & R (km) & $R_{1.4}$ (km) & $\Lambda_{1.4}$     \\
\hline
200  & 0.000  & 2.41 & 11.86 & 11.37 &  644 \\
200  & 0.02  & 2.37 & 11.75 & 11.22 &  586  \\
200  & 0.04  & 2.16 & 10.70  &10.39 &  346  \\
200  & 0.06  & 1.80 & 8.72 & 8.95 & 108     \\
\hline
60 &  0.000  & 2.41 & 11.86 & 11.37 & 644   \\
60  & 0.02   &2.40 & 11.84 & 11.30 & 625    \\
60  & 0.04  & 2.33 & 11.46 & 11.05 & 524    \\
60  & 0.06  & 2.16 & 11.31 & 10.42 & 351    \\
\hline \hline
\end{tabular}
\label{T2}
\end{table}
\subsection*{2. Fermionic DM within CFL quark matter}
We now study the effect of Fermionic DM in CFL superconducting matter described by the analytical approximation of Eq.~\ref{SCEOS}.
\begin{figure}
\centering
\includegraphics[width=0.32\textwidth,angle=270]{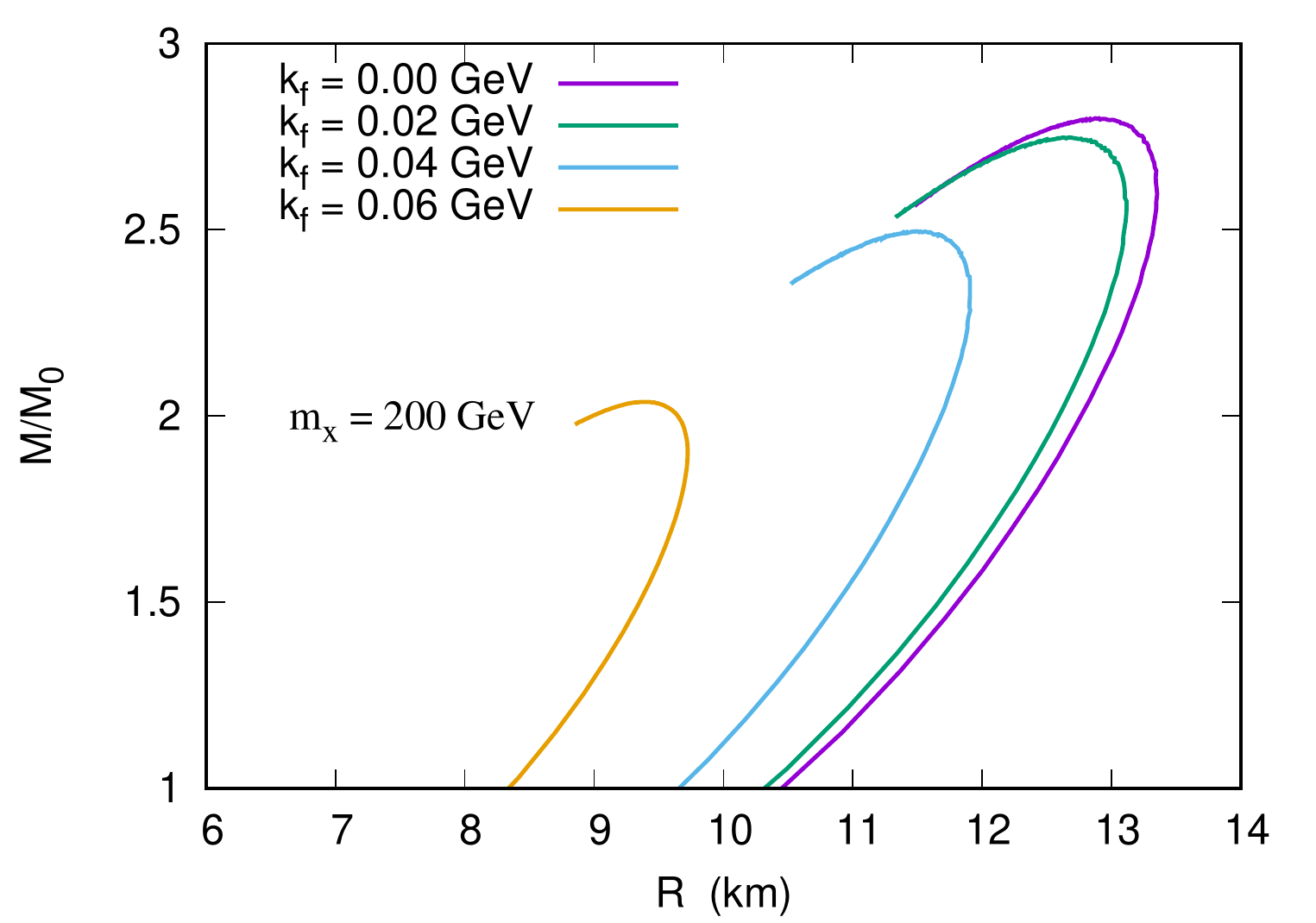} 
\includegraphics[width=0.32\textwidth,,angle=270]{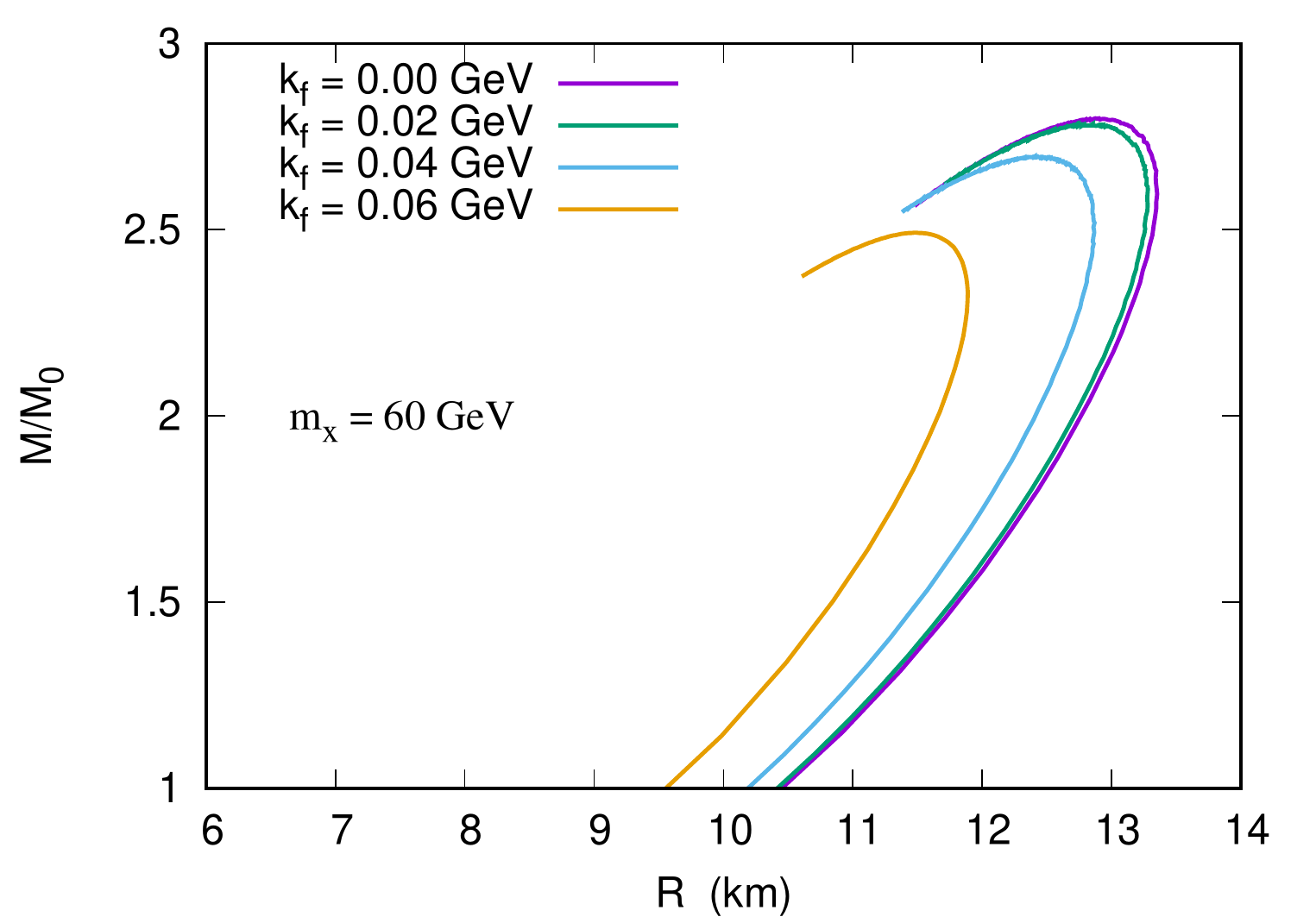}
\caption{Mass-radius relation for fermionic DM admixed CFL superconducting strange stars with $m_x =200$ GeV (left) and $m_x = 60$ GeV (right).}
\label{F3SC}
\end{figure}

We display in Fig.~\ref{F3SC} the TOV solution for a fermionic DM with a mass of 200 GeV and 60 GeV. As in the case of the vector MIT bag model, for CFL superconducting quark matter, the results for fermionic DM are significantly different when compared with bosonic DM. And again, the qualitative effect of fermionic DM is the same for CFL as it is for the vector MIT bag model. The maximum masses are always reduced, and the star compression always increases, even for very low masses. Again, different DM fractions always produce different mass-radius relations, affecting all the strange star families.

From the quantitative point of view, the maximum mass drops from 2.81 $M_\odot$ for $k_f^{\rm DM}$ = 0.00 to 2.04 $M_\odot$ for $k_f^{\rm DM}$ = 0.06 GeV in the case of $m_x$ = 200 GeV and to 2.49 $M_\odot$ for $m_x$ = 60 GeV. In the same sense, the radius of the canonical star drops from 11.57 km for $k_f^{\rm DM}$ = 0.00 to  9.21 km for $k_f^{\rm DM}$ = 0.06 GeV in the case of $m_x$ = 200 GeV, and 10.66 km for $m_x$ = 60 GeV. 
Now, unlike the case of the vector MIT, none of the CFL superconducting strange stars can be ruled out in the light of the  PSR J0740+6620, $M~=$ 2.08 $\pm$ 0.07 $M_\odot$~\cite{Fonseca_2021}, although for $k_f^{\rm DM}$ = 0.06 GeV and  $m_x$ = 200 GeV the radius of the canonical star is below the lower limit of 10.1 km presented in Ref.~\cite{Ozel_2016}.

\begin{figure}
\centering
\includegraphics[width=0.333\textwidth,angle=270]{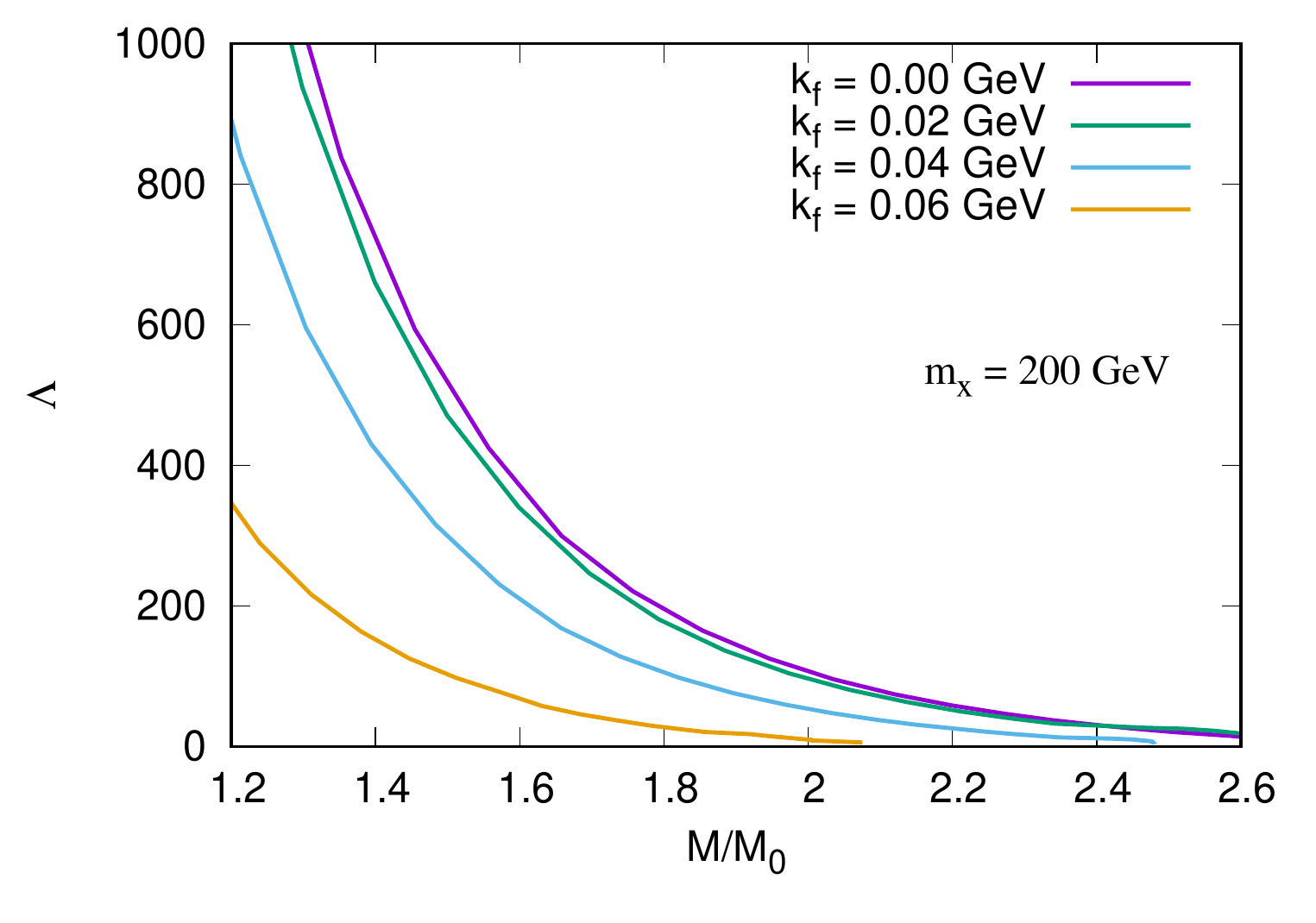}
\includegraphics[width=0.333\textwidth,angle=270]{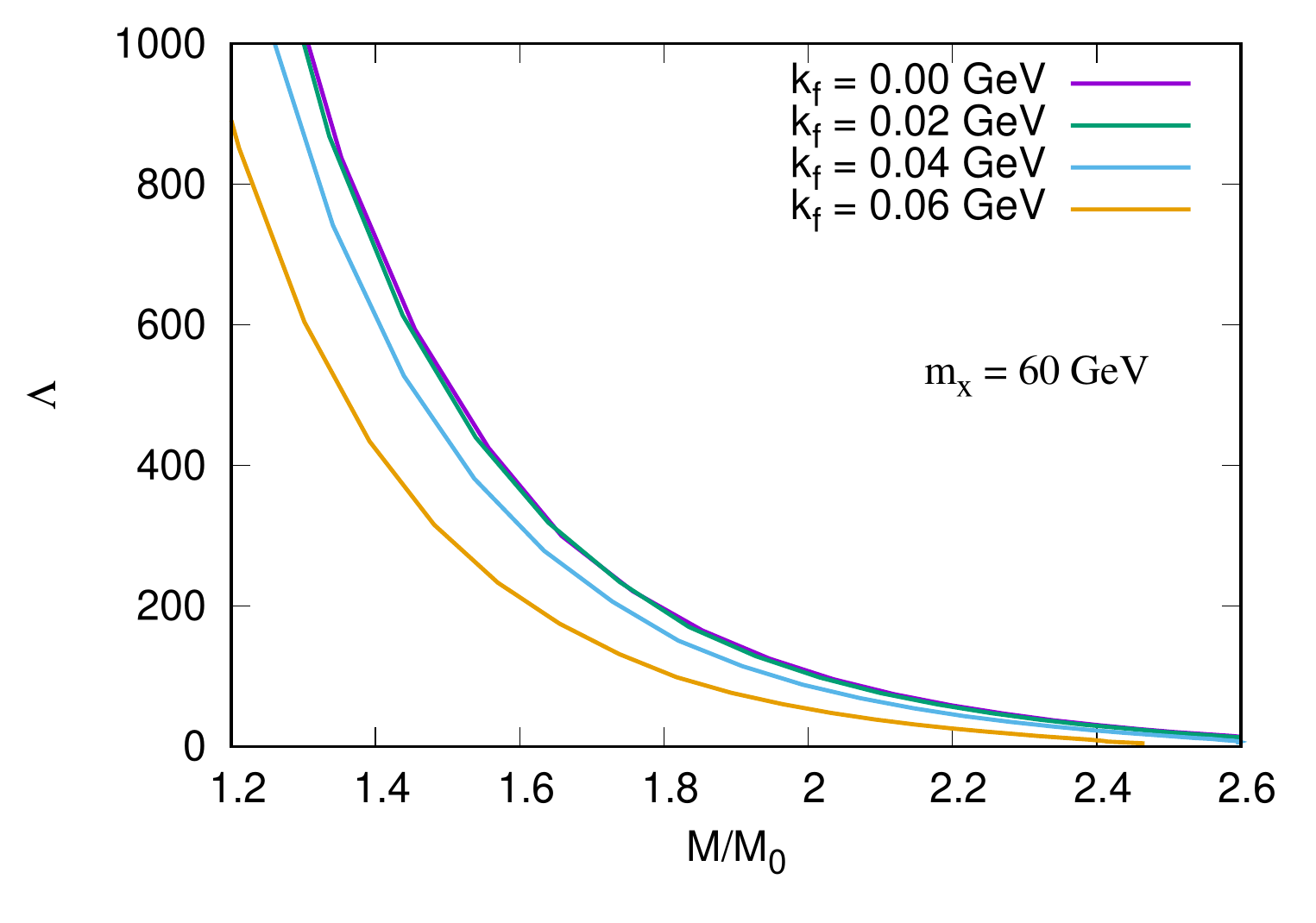}
\caption{Dimensionless tidal parameter $\Lambda$ for fermionic DM admixed CFL strange stars with $m_x=200$ GeV (left) and $m_x = 60$ GeV (right).} 
\label{F4SC}
\end{figure}

We plot in Fig.~\ref{F4SC} the dimensionless parameter $\Lambda$ for fermionic DM admixed superconducting strange stars with $m_x$ = 200 GeV and $m_x$ = 60 GeV. The results are completely analogous to the case of the vector MIT bag model; however, the value of $\Lambda$ here is always higher. The compression due to the fermionic DM contribution reduces the tidal parameter. In the case with $m_x$ = 200 GeV and $k_f^{\rm DM}$ = 0.06 GeV, the tidal parameter drops from 721 to 151. 

It is also worth noting that some parametrizations can fulfill the main constraints for pulsar observations,  70 $< \Lambda < 580$, and yet produce a very high maximum mass, sometimes reaching 2.50 $M_\odot$. The presence of DM again improves the theoretical prediction and the observational constraints, although it can be some debate about the radius of the canonical star. They do not fulfill NICER results~\cite{Riley_2019, Miller_2019}, but agree with Ref.~\cite{Capano_2020}.  Moreover, most parametrizations can explain even the black widow pulsar PSR J0952-0607~\cite{Romani_2022}.

Finally, even when we use a different model for the quark matter, the existence of almost degenerate results is still present: for $m_x$ = 200 GeV with $k_f^{\rm DM}$ = 0.04 GeV and $m_x$ = 60 GeV with $k_f^{\rm DM}$ = 0.06 GeV. The main results are summarized in Tab.~\ref{T2SC}.
\begin{table}
\centering
\caption{Macroscopic properties of fermionic DM admixed color  superconducting quark stars}
\begin{tabular}{ccccccc}
\hline \hline
$m_x$ (GeV)  &  $k_f^{\rm DM}$ (GeV) & $M/M_\odot$ & R (km) & $R_{1.4}$ (km) & $\Lambda_{1.4}$     \\
\hline
200  & 0.000  & 2.81 & 12.89 & 11.57 &  721 \\
200  & 0.02  & 2.75 & 12.65 & 11.43 &  653  \\
200  & 0.04  & 2.50 & 11.50  &10.67 &  421  \\
200  & 0.06  & 2.04 & 9.38 & 9.21 & 151     \\
\hline
60 &  0.000  & 2.81 & 12.89 & 11.57 & 721   \\
60  & 0.02   &2.78 & 12.75 & 11.53 & 694    \\
60  & 0.04  & 2.69 & 12.45 & 11.28 & 610    \\
60  & 0.06  & 2.49 & 11.53 & 10.66 & 422   \\
\hline \hline
\end{tabular}
\label{T2SC}
\end{table}
\subsection*{3. Fermionic DM with a vector channel}
Now we study if the presence of a dark, repulsive vector channel affects the macroscopic properties of the fermionic DM admixed strange stars. The new Lagrangian is  the Lagrangian in Eq.~\ref{FDMEOS} plus the repulsive channel and the respective meson mass, and reads~\cite{Guha_2021}:
\begin{equation}
\mathcal{L}_{\rm VDM} =  g_{\xi}\bar{\chi}(\gamma^\mu\xi_\mu)\chi + \frac{1}{2}m_\xi^2\xi_\mu\xi^\mu - \frac{1}{4}V^{\mu\nu}V_{\mu\nu} . \label{EV}
\end{equation}
The Lagrangian of Eq.~\ref{EV} is analogous to the $\omega$ contribution to the QHD Lagrangian~\cite{Lopes_2021, Serot_1992}. Indeed, the junction of Eq.~\ref{FDMEOS} and Eq.~\ref{EV} makes this model of DM fully analogous to the original $\sigma-\omega$ model of the QHD~\cite{Serot_1992}. The coupling constant $g_{\xi}=0.1$ is fixed, and it is equal to the $g_H$, while the mass of the vector of the dark meson is assumed to be 34 MeV, following Ref.~\cite{Guha_2021}. As the mass of the dark vector meson is 3000 times smaller than the mass of the Higgs boson, the repulsive channel is much stronger than the attractive one. Indeed, we have
\begin{equation}
G_\xi = \bigg ( \frac{g_\xi}{m_\xi} \bigg )^2 = 0.337 \quad \mbox{fm}^{2},
\end{equation} 
which is stronger than the quark repulsion and millions of times higher than the DM scalar coupling (Eq.~\ref{higgsc}). Nevertheless, despite the strong self-repulsion of the fermionic DM, the numerical results are barely affected by the repulsive channel. For the vector MIT bag model, the only noticeable difference appears for $m_x$ = 200 GeV and $k_f^{\rm DM}$ = 0.04 GeV. In this case, the maximum mass increase from 2.16 $M_\odot$ to 2.17 $M_\odot$. The radius of the canonical star also grows from 11.39 km to 11.46 km. The tidal parameter $\Lambda_{1.4}$ also increases from 346 to 358. It is worth noticing that all these variations are far beyond the precision with which experimental measurements are made.   All the other parametrizations present even lower (or none) differences. Herefore, we do not provide any figures in this section since they would be visually indistinguishable from those in the last paragraph. We only display the main results in Tab.~\ref{T3}. In the case of CFL superconducting quark matter, the differences are even smaller.

The nature of the vector coupling can explain why the differences are so small. The vector mesons couple to the number density, and we are dealing with a very low-density regime. Indeed, even $k_f^{\rm DM}$ = 0.06 GeV implies a number density is around 9.6 $\times~10^{-4}$ fm$^{-3}$. Of course, we could increase the repulsion of the dark vector boson, but we believe this would be very unrealistic since DM was proposed to explain higher attraction in galaxy curves~\cite{Franco}.

\begin{table}
\centering
\caption{Macroscopic properties of dark vector boson fermionic DM admixed strange stars within the vector MIT bag model. The only significant differences are for $k_f^{\rm DM}$ = 0.04 GeV }
\begin{tabular}{ccccccc}
\hline \hline
$m_x$ (GeV)  &  $k_f^{\rm DM}$ (GeV) & $M/M_\odot$ & $R$ (km) & $R_{1.4}$ (km) & $\Lambda_{1.4}$     \\
\hline
200  & 0.000  & 2.41 & 11.86 & 11.37 &  644  \\
200  & 0.02  & 2.37 & 11.75 & 11.22 &  586   \\
200  & 0.04  & 2.17 & 10.70  &10.46 &  358   \\
200  & 0.06  & 1.80 & 8.72 & 9.01 & 112      \\
\hline
60 &  0.000  & 2.41 & 11.86 & 11.37 & 644    \\
60  & 0.02   &2.40 & 11.84 & 11.30 & 625     \\
60  & 0.04  & 2.33 & 11.46 & 11.10 & 532     \\
60  & 0.06  & 2.16 & 11.31 & 10.43 & 353     \\
\hline \hline
\end{tabular}
\label{T3}
\end{table}
\section{Conclusions}
In this work, we calculate the properties for the DM admixed for strange quark stars. We use two different models for the quark model:  the vector MIT bag model, as presented in Refs.~\cite{Lopes_2021,Lopes_2022} and the CFL color superconducting quark matter via an analytical approximation, as discussed in Refs.~\cite{Zdunik2013,Han2019,Alford2013}; and two different kinds of dark matter: a bosonic as discussed in Refs.~\cite{Li_2012_1,Li_2012_2,Panotopoulos_2017,Lopes_2018} and for fermionic ~\cite{OdilonDM,Das_2021,Lourenco_2021}. For each kind of DM, we use two different mass values, and the strange stars always agree with the  Bodmer-Witten conjecture \cite{Bodmer_1971, Witten_1984}. Our main conclusions can be summarized as follows:
\begin{itemize}

    \item The qualitative results for DM admixed strange stars are independent of the quark model utilized. This is true for both bosonic and fermionic, as well it is independent of the DM mass.

    \item For a bosonic DM with a mass of $m_x$ = 100 MeV, we have an increase of the maximum mass, while the properties of low-mass strange stars are not significantly affected. This is the only case in that we have an increase in the star's mass. Such a situation happens for the vector MIT, the  CFL superconducting quark matter, and also for the massless MIT, as pointed out in Refs.~\cite{Lopes_2018, Panotopoulos_2017}.

    \item For a bosonic DM with a mass of $m_x$ = 400 MeV, we have a decrease of the maximum mass, whilst the radii of the low-mass strange stars, in this case, are also affected.

    \item For a fermionic DM, the maximum mass always decreases. The higher the DM fraction, the lower the maximum mass, and the smaller the radii. Also, the higher the DM mass, the higher the stellar compression and the lower the maximum mass.

    \item Although we introduce a repulsive dark vector field with a mass 3000 times smaller than the attractive scalar field, we do not find significant variation in the stellar macroscopic properties.

    \item There are almost degenerate results both for $m_x$ = 200 GeV with $k_f^{\rm DM}$ = 0.04 GeV and $m_x$ = 60 GeV with $k_f^{\rm DM}$ = 0.06 GeV, the maximum mass, as well the properties of the canonical star are essentially the same.

    \item About the observational constraints, we can see that the mass of the PSR J0740+6620 pulsar, $M~=$ 2.08 $\pm$ 0.07 $M_\odot$~\cite{Fonseca_2021} is easily obtained. Even the mass range of  2.35 $\pm$ 0.17 $M_\odot$ of the black widow pulsar PSR J0952-0607~\cite{Romani_2022} can be reached for some parametrization. 

    \item The radius of the canonical star is still a matter of debate. Most of our results point to a radius between 11.0 km to 11.5 km. In general, our results are in agreement with Ref.~\cite{Capano_2020} but are too low to reproduce the NICER results~\cite{Riley_2019, Miller_2019} whilst at the same time are too high to agree with Ref.~\cite{Ozel_2016}.

   \item Except for bosonic DM with a mass of $m_x$ = 100 MeV, in all other cases, the presence of the DM reduces the dimensionless tidal parameter $\Lambda$. In most of these cases, the constraint  70 $< \Lambda < 580$~\cite{Abbott_2018} is easily fulfilled. 
\end{itemize}

\appendix

\section{The Tidal deformability parameter}

Let us start with the well known  TOV equations~\citep{Oppenheimer_1939}:

\begin{eqnarray}
 \frac{dp}{dr} &=& \frac{-GM(r)\epsilon (r)}{r^{2}} \bigg [ 1 + \frac{p(r}{\epsilon(r)} \bigg ] \times \bigg  [ 1 + \frac{4\pi p(r)r^3}{M(r)} \bigg ] \bigg [ 1 - \frac{2GM(r)}{r} \bigg ]^{-1}, \nonumber \\
 \frac{dM}{dr} &=&  4\pi r^2 \epsilon(r) . \label{EL11}
\end{eqnarray}

The dimensionless
tidal deformability parameter $\Lambda$ is defined as:

\begin{equation}
 \Lambda~\equiv~\frac{\lambda}{M^5}~\equiv~\frac{2k_2}{3C^5} , \label{EL13}
\end{equation}
where $M$ is the compact object mass and $C = GM/R$ is its
compactness. The parameter $k_2$ is called the second (order) Love number:

\begin{eqnarray}
 k_2 &=&  \frac{8C^5}{5}(1-2C)^2 \times [2 + 2C(y_R -1) -y_R] 
 \{ 2C[6 -3y_R +3C(5y_R -8)] \nonumber \\
 &+&4C^3[13 -11y_R +C(3y_R -2) +2C^2(1+y_R)]+3(1-2C)^2 \nonumber \\
 &&[2-y_R +2C(y_R -1))]\ln(1-2C)\} ^{-1}, \nonumber
 \\ \label{EL14}
\end{eqnarray}
where $y_R=y(r=R)$ and $y(r)$ is obtained  by solving:

\begin{equation}
 r\frac{dy}{dr} +y^2 + yF(r) +r^2Q(r) = 0 . \label{EL15}
\end{equation}
 Eq.~(\ref{EL15}) must be solved coupled with the TOV equations,
 Eq.~(\ref{EL11}). The coefficients $F(r)$ and $Q(r)$ are given by:

\begin{equation}
 F(r) = \frac{1 -4\pi G r^2[\epsilon(r) - p(r)]}{E(r)} ,
\end{equation}

\begin{eqnarray}
 Q(r) = \frac{4\pi G}{E(r)} \bigg [ 5\epsilon(r) + 9p(r) + \frac{\epsilon(r)+p(r)}{\partial p/ \partial \epsilon} - \frac{6}{4\pi Gr^2} \bigg ]  
 -4 \bigg [ \frac{G[M(r) + 4\pi r^3 p(r)]}{r^2 E(r)} \bigg ]^2, \label{EL17}
\end{eqnarray}
where $E(r) = (1 -2GM(r)/r)$.  Additional discussion about the theory of tidal deformability and the tidal Love numbers are beyond the scope of this work and can be found in Refs.~\cite{Abbott_2017, Abbott_2018, Flores_2020, Chatziioannou_2020, Postnikov_2010, Lourenco_2021, Lopes_ApJ,Lenzi2023} and references therein.
\bibliography{dm}
\bibliographystyle{JHEP}
\end{document}